\documentclass{article}
\usepackage[margin=0.8in]{geometry}
\usepackage{setspace,url,hyperref,graphicx}
\usepackage{authblk}
\usepackage{todonotes}
\usepackage{amsmath,amsthm,amssymb,amsfonts}
\newcommand{\N}{\ensuremath{\mathbb{N}}}
\newcommand{\R}{\ensuremath{\mathbb{R}}}

\usepackage[sorting = none, backend = bibtex, style=numeric-comp]{biblatex}
\title{Posiform Planting: Generating QUBO Instances for Benchmarking}

\author[1]{Georg Hahn\thanks{Email: ghahn@hsph.harvard.edu}}
\author[2]{Elijah Pelofske\thanks{Email: epelofske@lanl.gov}}
\author[2,3]{Hristo N.\ Djidjev}
\affil[1]{Harvard University, T.H.\ Chan School of Public Health}
\affil[2]{Los Alamos National Laboratory, CCS-3 Information Sciences}
\affil[3]{Bulgarian Academy of Sciences, Institute of Information and Communication Technologies}
\date{\vspace{-0.8cm}}

\begin{document}
\maketitle

\begin{abstract}
We are interested in benchmarking both quantum annealing and classical algorithms for minimizing Quadratic Unconstrained Binary Optimization (QUBO) problems. Such problems are NP-hard in general, implying that the exact minima of randomly generated instances are hard to find and thus typically unknown. While brute forcing smaller instances is possible, such instances are typically not interesting due to being too easy for both quantum and classical algorithms. In this contribution, we propose a novel method, called \textit{posiform planting}, for generating random QUBO instances of arbitrary size with known optimal solutions, and use those instances to benchmark the sampling quality of four D-Wave quantum annealers utilizing different interconnection structures (Chimera, Pegasus, and Zephyr hardware graphs) as well as the simulated annealing algorithm. Posiform planting differs from many existing methods in two key ways. It ensures the uniqueness of the planted optimal solution, thus avoiding groundstate degeneracy, and it enables the generation of QUBOs that are tailored to a given hardware connectivity structure, provided that the connectivity is not too sparse. Posiform planted QUBOs are a type of 2-SAT boolean satisfiability combinatorial optimization problems. Our experiments demonstrate the capability of the D-Wave quantum annealers to sample the optimal planted solution of combinatorial optimization problems with up to $5627$ qubits.
\end{abstract}

\section{Introduction}
\label{sec:introduction}
Many important NP-hard optimization problems can be easily expressed in a QUBO (quadratic unconstrained binary optimization) or an Ising form \cite{Lucas2014}, given by the quadratic function 
\begin{align}
    Q(x_1,\ldots,x_n) = \sum_{i=1}^n a_i x_i + \sum_{i<j} a_{ij} x_i x_j
    \label{eq:QUBO_form}
\end{align}
in $n \in \N$ binary variables. In eq.~\eqref{eq:QUBO_form}, the linear weights $a_i \in \R$ and the quadratic couplers $a_{ij} \in \R$ define the problem under investigation and are chosen by the user. The assignments of the variables $x_i$ for $i \in \{1,\ldots,n\}$ are unknown, and we seek a configuration of $(x_1,\ldots,x_n)$ minimizing eq.~\eqref{eq:QUBO_form}. If $x_i \in \{0,1\}$ then eq.~\eqref{eq:QUBO_form} is called a QUBO problem, and if $x_i \in \{-1,+1\}$ it is called an Ising model.

Since many NP-Hard problems can be formulated as QUBO models, it is of interest to efficiently compute the optimal solution(s) of general QUBO problems. To this end, researchers have developed a variety of classical approaches \cite{Boros2006, Boros2007, kirkpatrick1983optimization} to compute solutions of high quality that minimize eq.~\eqref{eq:QUBO_form}. Quantum annealing offers an experimental route to sample combinatorial problems. Quantum annealing is a type of analog quantum computation that uses quantum fluctuations to attempt to arrive at an optimal (or a very good) minimum of eq.~\eqref{eq:QUBO_form} \cite{Kadowaki_1998, das2008colloquium, morita2008mathematical, Hauke_2020}. The quantum annealing algorithm has been physically instantiated in a number of ways, including superconducting flux qubit hardware that is manufactured by D-Wave Systems, Inc. The D-Wave quantum annealers have been evaluated for sampling a large number of different types of problems, typically focusing on combinatorial optimization problems or Hamiltonian dynamics \cite{Lanting_2014, king2021scaling, Boixo_2014, King_2023, King_2022, boixo2016computational, harris2018phase, Venturelli_2015, boixo2013experimental, tasseff2022emerging}. D-Wave quantum annealing devices offer on the scale of hundreds to thousands of qubits, but are still subject to connectivity constraints, control errors, and noise from the environment \cite{Zaborniak_2021, Pelofske_2023, pearson2019analog, grant2022benchmarking, nelson2021singlequbit, lanting2020probing}. In order to map a QUBO $Q$ of eq.~\eqref{eq:QUBO_form} directly on the hardware chip of a quantum annealer, its connectivity structure should be consistent with the connectivity structure of the quantum device. Specifically, each variable $x_i$ is mapped to a distinct qubit $q_i$. For each nonzero coefficient $a_{ij}$, there should be a coupler (direct link) between qubits $q_i$ and $q_j$. If a direct embedding is not possible, then a \textit{minor embedding} of the graph representing the sparsity structure of the QUBO $Q$ onto the graph defined by the hardware structure can be used \cite{PRXQuantum.2.040322, PhysRevA.105.022615, choi2008minor, choi2011minor}. However, the number of qubits required in that case may grow quadratically with the size of $Q$.

To better assess the capabilities of both classical and quantum approaches for sampling (approximate) solutions of combinatorial optimization problems, methods are needed that generate benchmark problems with (ideally) known solutions. Two strategies exist to achieve this goal. First, one can generate problems of the type of eq.~\eqref{eq:QUBO_form} with randomly sampled linear and quadratic weights, and then brute force them. However, brute forcing is only feasible for problems with a relatively small number of variables (roughly $30$ variables for full brute force computations). Second, methods have been developed that allow one to generate QUBO problems with planted solutions, that is, problems generated to have a solution that is specified a-priori. A detailed overview of such methods is given in Section~\ref{sec:literature}. Importantly, existing methods often have two shortcomings. Many approaches only ensure that the generated problem has a minimum at the planted solution, but do not guarantee its uniqueness. Moreover, for many methods the sparsity structure of the generated QUBO cannot be chosen, which means the QUBO cannot directly be solved on certain hardware devices. Naturally, since the minimization of eq.~\eqref{eq:QUBO_form} is NP-hard, all methods exploit some form of shortcut or mathematical device to generate large problems with nontrivial structures and known solutions.

In this contribution, we introduce a new method to generate QUBO problems of the type of eq.~\eqref{eq:QUBO_form} with a single planted solution. The method is called \textit{posiform planting}, in reference to the mechanism we exploit that generates a QUBO in posiform representation. The posiforms are converted to QUBOs only at a later stage when the solution has been planted. Two features of our algorithm are noteworthy. First, it guarantees the uniqueness of the planted solution. Moreover, the connectivity structure of the QUBO can, in principle, be chosen freely. Naturally, the generated QUBOs need to have at least a certain number of quadratic terms to guarantee the uniqueness of the planted solution and thus cannot be too sparse, although this also depends on the solution being planted. In contrast to some existing solution-planting methods, such as the \textit{tile planting} or \textit{deceptive cluster loops} methods of the \textit{Chook} toolbox \cite{chook}, posiform planting generates QUBO problems which include linear terms.

The adaptation to an arbitrary connectivity structure is of importance when generating problems that are tailored to, for instance, the qubit connectivity structure of the D-Wave quantum annealers. In particular, the physical qubits across currently existing D-Wave generations use connections determined by Chimera, Pegasus, or Zephyr graphs \cite{dattani2019pegasus, boothby2020nextgeneration}. Being able to tailor the generated problems to any arbitrary architecture allows one to generate much larger benchmark problems compared to the case where the problems cannot be directly embedded, thus necessitating the computation of a minor embedding onto the D-Wave QPU chip structure.

One of the properties of the transverse field driver in quantum annealing and other approximate quantum optimization algorithms is that degenerate ground states are not in general sampled uniformly \cite{Zhang_2017, 9605329, matsuda2009quantum, PhysRevA.100.030303, kumar2020achieving, Nelson_2022, Mandra2017, konz2019embedding}. Posiform planting guarantees the uniqueness of the planted optimal solution, thus any use cases in which biased sampling of degenerate solutions needs to be avoided could benefit from posiform planting. Some use cases in which biased sampling of degenerate solutions should be avoided include the estimation the ground-state entropy of a degenerate physical systems, estimating the count of the total number of solutions in combinatorics, or the estimation of ground state probabilities in industrial applications where the problem has several solutions by design \cite{Mandra2017}.

This article is structured as follows. After a literature review in Section~\ref{sec:literature}, we introduce the idea of posiform planting in Section~\ref{sec:methods}. We evaluate the QUBO problems generated by posiform planting on D-Wave devices using both native connectivity (using the Chimera, Pegasus, and Zephyr hardware graphs), as well as arbitrarily connected minor embedded problem instances (Section~\ref{sec:results}). The hardware native QUBOs are also sampled using the classical heuristics simulated annealing and steepest gradient descent. The article concludes with a discussion in Section~\ref{sec:discussion}. Data and extra figures generated from this research are publicly available as a Zenodo dataset \cite{georg_hahn_2023_8336707}.

\subsection{Literature review}
\label{sec:literature}
A variety of contributions in the literature focus on the generation of QUBO or Ising models of the type of eq.~\eqref{eq:QUBO_form} that can serve as benchmark problems. These methods can be grouped according to the underlying mechanism they use to generate problems, and according to the properties they guarantee. Originally this property of known planted solutions was introduced from satisfiability problems \cite{Barthel_2002, Krzakala_2009}.

One popular way to generate problems is with the help of frustrated loops, meaning Ising models of the form $Q=\sum_{j=1}^M Q_j$, where each $Q_j$ only contains a subset of the variables. For instance, \cite{Hen2015, king2015performance} generate frustrated Ising models with tunable hardness, though the authors explicitly point out that they cannot guarantee uniqueness. Similar methods are the so-called tile-planting \cite{Perera2020} and patch-planting for Ising models \cite{Wang2017}. In \cite{Pei2020} the authors generate weighted MAX-2-SAT instances with the help of frustrated loops that have known solutions. Notably, the hardness of their problems can be tuned through a parameter called the frustration index.

One major drawback of many published planted solution methods is the fact that they do not guarantee the uniqueness of the planted solution, meaning that the input configuration is only guaranteed to be one of a possibly unknown number of minima. A notable exception is \cite{Kowalsky2022}, who ensure the uniqueness of solution with an approach based on equation planting. However, the resulting QUBOs have a very special form as each linear equation is required to contain exactly three binary variables.

Another route is called equation planting, that is the generation of QUBO problems from a set of (linear) equations. In \cite{Hen2019}, the author considers a set of linear equations modulo $2$ to pin down the bitstring to be planted, and then recasts it as an Ising model. Their method is based on the experimental observation that although linear equations are easy to solve, they disguise the solution well for machines when being recast as an optimization problem. According to the author, equation planting guarantees the uniqueness of the planted solution. However, tailoring the instances to a given connectivity structure is not mentioned.

A popular tool for generating binary optimization problems with planted solutions is the \textit{Chook} toolbox of \cite{chook}. Chook implements several approaches, named "tile planting", "Wishart planting", "equation planting", and "$k$-local planting". However, none of those approaches guarantees uniqueness, and some of them (such as Wishart planting) are not designed to tailor to arbitrary connectivity structures. Notably, the method "deceptive cluster loops" is tailored to the D-Wave Chimera topology.

The software package \textit{dwig} contains Python implementations of several existing planted solution methods, specifically, \textit{RAN-pr} \cite{zdeborova2016statistical}, \textit{RAN-k} \cite{king2015benchmarking}, \textit{FL-k} \cite{king2015performance}, \textit{FCL-k} \cite{king2017quantum}, \textit{weak-strong cluster network} \cite{denchev2016computational}, \textit{frustrated cluster loops} \cite{albash2018advantage}, and \textit{corrupted biased ferromagnet} \cite{pang2020structure}.

There are several studies which have examined the sampling of MAX 2-SAT combinatorial optimization problems using quantum annealing, some with an emphasis on generating MAX 2-SAT, which are challenging for quantum annealing to sample \cite{crosson2014different, Mehta_2022, PhysRevResearch.5.023151, Mehta_2021, hsu2018quantum, Santra_2014}.

The above methods have been used in a number of studies on sampling characteristics of quantum annealers \cite{PhysRevApplied.11.044083, Zhang_2017, Barash_2019, king2015benchmarking}.

\section{Methods}
\label{sec:methods}
This section introduces a novel method to generate QUBO models of the type of eq.~\eqref{eq:QUBO_form} for a customized connectivity structure and a with a guarantee of uniqueness for the planted solution. The method is based on the generation of a posiform representation of eq.~\eqref{eq:QUBO_form}, which is introduced in Section~\ref{sec:posiform}. The construction of the posiform and the guarantee of uniqueness are based on the fact that testing if a posiform attains the value zero is equivalent to a 2-SAT problem, which can be solved in polynomial time (Section~\ref{sec:2sat}). The complete algorithm is summarized in Section~\ref{sec:generation}. A note on how the generation can naturally be adapted to a given connectivity structure is discussed in Section~\ref{sec:connectivity}.

\subsection{Conversion from QUBO to posiform}
\label{sec:posiform}
A \textit{posiform} is a quadratic function with positive coefficients on an extended set of variables $\mathcal{Z} = \{ x_1, \ldots, x_n \} \cup \{ \overline{x}_1, \ldots, \overline{x}_n \}$, meaning that a posiform can contain either a variable $x_i \in \{0,1\}$ or its complement $\overline{x}_i = 1-x_i$, where $i \in \{ 1,\ldots,n \}$. A posiform can be expressed as
\begin{align}
    P(x_1,\ldots,x_n)=P(x_1,\ldots,x_n,\overline{x}_1,\ldots,\overline{x}_n) = \sum_{z \in \mathcal{Z}} b_z z + \sum_{z,z' \in \mathcal{Z}} b_{z z'} z z',
    \label{eq:posiform}
\end{align}
where each $z \in \mathcal{Z}$ and $z' \in \mathcal{Z}$ stand for one of the variables $x_i$ or its complement $\overline{x}_i$, $i \in \{ 1,\ldots,n \}$ and the coefficients $b_z$ and $b_{z z'}$ are nonnegative.

Any QUBO of the form of eq.~\eqref{eq:QUBO_form} can be written as a posiform. To this end, consider first the linear terms. If $a_i>0$ for some $i \in \{ 1,\ldots,n \}$ in eq.~\eqref{eq:QUBO_form}, it remains unchanged in the posiform. If $a_i<0$, we rewrite $a_i x_i = a_i (1-\overline{x}_i) = a_i + (-a_i) \overline{x}_i$. The single summand $a_i$ is constant and can be omitted as it does not impact the location of the minimum of eq.~\eqref{eq:QUBO_form}. The term $(-a_i) \overline{x}_i$ complies with the posiform requirement as $-a_i>0$ given $a_i<0$.

Similarly, any quadratic term $a_{ij} x_i x_j$ with $a_{ij}>0$ in eq.~\eqref{eq:QUBO_form} remains unchanged in the posiform. If $a_{ij} x_i x_j$ with $a_{ij}<0$ in eq.~\eqref{eq:QUBO_form}, we rewrite it as either $a_{ij} (1-\overline{x}_i) x_j = a_{ij} + (-a_{ij}) \overline{x}_j + (-a_{ij}) \overline{x}_i x_j$ or $a_{ij} x_i (1-\overline{x}_j) = a_{ij} + (-a_{ij}) \overline{x}_i + (-a_{ij}) x_i \overline{x}_j$. Both options are valid choices and none is preferable over the other. As can be seen, apart from the constant term $a_{ij}$, which can be omitted, the remaining summands have positive coefficients $-a_{ij}>0$ given $a_{ij}<0$.

As a simple example, consider the following QUBO in three variables, $Q(x_1,x_2,x_3) = 2 x_1 - x_2 + x_1 x_2 -2 x_2 x_3$. In posiform representation, it can be written as $P(x_1,x_2,x_3) = 2x_1 + \overline{x}_2 + 2 \overline{x}_3 + x_1 x_2 + 2 \overline{x}_2 x_3$, where we omitted the offset $-3$ that results from the conversion.

\subsection{Connection to 2-SAT problems}
\label{sec:2sat}
The idea of posiform planting is to generate posiforms that attain a value zero at a unique known (planted) combination of  values of the variables. Assume a posiform of the type of eq.~\eqref{eq:posiform} is given. Clearly the minimum of eq.~\eqref{eq:posiform} is bounded below by zero as all coefficients and variables are nonnegative. Moreover, we can test if there is a configuration $x=(x_1,\ldots,x_n)$ that achieves $P(x_1,\ldots,x_n)=0$ in linear time.

This can be seen as follows. If $P(x_1,\ldots,x_n)=0$, then all summands in eq.~\eqref{eq:posiform} must be zero. Therefore, we aim to find $x=(x_1,\ldots,x_n)$ such that $z=0$ for all linear terms, and $z z' = 0$ for all quadratic terms in eq.~\eqref{eq:posiform}, where $z,z' \in \mathcal{Z}$. For the quadratic terms, $z z' = 0$ is equivalent to $\overline{z} \vee \overline{z}'=\textit{True}$. We thus rewrite all linear and quadratic terms in eq.~\eqref{eq:posiform} without their coefficients into a 2-SAT problem, which can be solved in linear time \cite{Krom1967, Even1976, Aspvall1979}. Any solution to the constructed 2-SAT problem will satisfy $P(x_1,\ldots,x_n)=0$ and vice versa. Importantly, if the 2-SAT problem has a unique solution, so does the corresponding posiform.

\subsection{QUBO generation with given connectivity and planted unique solution}
\label{sec:generation}
We are given a bitstring $x^\ast = (x_1^\ast,\ldots,x_n^\ast)$ denoting the solution to be planted. The first step is to generate a 2-SAT problem having $x^\ast$ as its unique solution. We aim to construct a 2-SAT problem having $x^\ast$ as its unique solution with the help of an exclusion argument, meaning that we add clauses to the 2-SAT problem that exclude any bitstring other than $x^\ast$. This is achieved as follows.

We select two random indices $i,j \in \{1,\ldots,n\}$ with $i \neq j$ and consider the two bits $x_i^\ast$ and $x_j^\ast$ in the solution to be planted. We then randomly select one of the three possible binary tuples $(\hat{x}_i,\hat{x}_j)$ satisfying $(\hat{x}_i,\hat{x}_j) \neq (x_i^\ast,x_j^\ast)$. Depending on the choice of $(\hat{x}_i,\hat{x}_j)$, we add a clause to the current 2-SAT problem that excludes the possibility of $(x_i,x_j)=(\hat{x}_i,\hat{x}_j)$ in an optimal solution, precisely, the clause
\begin{align}
\begin{split}
\neg (\overline{x}_i \wedge \overline{x}_j) &= (x_i \vee x_j) \qquad \text{if}~(\hat{x}_i,\hat{x}_j)=(0,0),\\
\neg (\overline{x}_i \wedge x_j) &= (x_i \vee \overline{x}_j) \qquad \text{if}~(\hat{x}_i,\hat{x}_j)=(0,1),\\
\neg (x_i \wedge \overline{x}_j) &= (\overline{x}_i \vee x_j) \qquad \text{if}~(\hat{x}_i,\hat{x}_j)=(1,0),\\
\neg (x_i \wedge x_j) &= (\overline{x}_i \vee \overline{x}_j) \qquad \text{if}~(\hat{x}_i,\hat{x}_j)=(1,1).
\end{split}
\label{eq:clauses}
\end{align}
After each added clause, we attempt to solve the generated 2-SAT problem at its current stage. By construction, the choice of the clauses added to the 2-SAT problem will never exclude the planted bitstring $x^\ast$ from the solution set of the generated 2-SAT problem.

We continue in this fashion until we arrive at a 2-SAT problem which has $x^\ast$ as its unique solution. Our procedure only requires polynomial effort. Indeed, it is known that the phase transition in 2-SAT problems occurs for $n$ variables at $O(n)$ clauses \cite{Gent1994, Coja2016}, thus we expect to only add a linear number of clauses until $x^\ast$ remains as the unique solution of the 2-SAT problem. Moreover, solving a 2-SAT problem can be done in linear time \cite{Krom1967, Even1976, Aspvall1979}. Note that, to save computational effort, it is not necessary to solve the 2-SAT problem being generated each time a new clause is added. Instead, it suffices to solve it after adding a certain batch size $B \in \N$ of new clauses. In the experiments of Section~\ref{sec:results}, we use the \textit{MiniSat} solver of \cite{minisat}.

Once a 2-SAT problem is constructed with $x^\ast$ as its unique solution, we construct a posiform from it. Thus, in the second step, we convert each clause $(z \vee z') = \neg (\overline{z} \wedge \overline{z}')$ into the quadratic term $b_{z z'} \overline{z}~\overline{z}'$, where $z, z' \in \mathcal{Z}$. The negation is necessary here as each clause $(z \vee z')$ that is \textit{True} (value 1) in the 2-SAT problem needs to be zero in the posiform (see Section~\ref{sec:2sat}) as it is a function to be minimized. Importantly, the coefficient $b_{z z'} > 0$ of the posiform is actually freely choosable (as long as it is positive). Substituting any complement $\overline{x}_i$ as $1-x_i$ and multiplying out the expression yields a QUBO with (typically) both positive and negative QUBO coefficients.

As an example, suppose we aim to plant the solution $x^\ast = (1,0,1)$ in $n=3$ variables. For the random indices $(i,j) = (2,3)$ we choose $(\hat{x}_2,\hat{x}_3) = (1,1)$, thus satisfying $(\hat{x}_2,\hat{x}_3) \neq (x_2^\ast,x_3^\ast)$. According to eq.~\eqref{eq:clauses}, we add the clause $(\overline{x}_2 \vee \overline{x}_3)$ to the 2-SAT problem being generated. By continuing in this fashion for other randomly chosen variable pairs in $x^\ast$, we might obtain the 2-SAT instance
\begin{align}
(\overline{x}_2 \vee \overline{x}_3) \wedge (x_1 \vee \overline{x}_2) \wedge (x_1 \vee \overline{x}_3) \wedge (x_1 \vee x_2) \wedge (\overline{x}_2 \vee x_3) \wedge (\overline{x}_1 \vee x_3),
\label{eq:example2sat}
\end{align}
which can easily be checked to have the unique solution $x^\ast$. Rewriting eq.~\eqref{eq:example2sat} into a posiform results in $P = x_2 x_3 + \overline{x}_1 x_2 + \overline{x}_1 x_3 + \overline{x}_1 \overline{x}_2 + x_2 \overline{x}_3 + x_1 \overline{x}_3$. Note that the coefficients of $P$ (set here to 1) can be freely chosen as long as they are positive. Multiplying out the posiform leads to the QUBO $Q(x_1,x_2,x_3) = x_2 + x_3 - 2 x_1 x_3$, which can easily be verified to have a unique minimum at $x^\ast$.

\subsection{Adaptation to connectivity structures}
\label{sec:connectivity}
Apart from the guarantee of uniqueness, the algorithm of Section~\ref{sec:generation} allows one to adapt the generated QUBOs to a given connectivity structure. This is possible since there are no restrictions on the choice of tuples $(x_i^\ast,x_j^\ast)$ with $i,j \in \{1,\ldots,n\}$ that are being used to narrow down the solution space to $x^\ast$ in the 2-SAT problem.

To be precise, instead of sampling $i,j \in \{1,\ldots,n\}$, it is valid to sample $(i,j) \in \mathcal{E}$ for some edge set $\mathcal{E} \subseteq \{1,\ldots,n\} \times \{1,\ldots,n\}$. When converting the generated 2-SAT problem to a posiform, the clauses become the quadratic terms, and when multiplying out the posiform into a QUBO, no further couplers are being introduced. Therefore, the edges in $\mathcal{E}$ will translate 1-to-1 to the quadratic couplers in the posiform and in the QUBO. For instance, $\mathcal{E}$ can be chosen as the fixed connectivity graph of one of the D-Wave annealer generations. Naturally, if $\mathcal{E}$ is too sparse, it might not be guaranteed any more that enough clauses can be sampled to narrow down $x^\ast$ as the unique solution, however this problem was not encountered for any of the D-Wave hardware graphs.

\section{Results}
\label{sec:results}
In this section, we investigate the performance of the posiform planting methodology introduced in Section~\ref{sec:methods}. The section starts with an overview of the D-Wave devices and their parameters in Section~\ref{sec:results_parameter_settings}. In Section~\ref{sec:results_native_QUBO}, we use posiform planting to generate and solve QUBO instances on four D-Wave machines that fit their hardware natively, thus allowing for very large instance sizes. The hardness of the generated instances is assessed by computing the ground state probability (GSP) as well as the time-to-solution (TTS) metrics. In Section~\ref{sec:results_minor_embedded}, we investigate instances with arbitrary qubit connectivity, thus requiring a minor embedding of the problem QUBO onto the D-Wave hardware.

\subsection{Parameter Settings}
\label{sec:results_parameter_settings}
Table~\ref{tab:hardware_summary} shows the four generations of the D-Wave quantum annealer used in the experiments of this section. Apart from the Chip ID and the name of the D-Wave topology, Table~\ref{tab:hardware_summary} displays the number of available qubits and couplers, and the annealing times supported by the device.

\begin{table}[h]
    \begin{center}
        \begin{tabular}{|l||l|l|l|l|}
            \hline
            D-Wave QPU Chip ID & Topology & Available & Available & Annealing time\\
            & name & qubits & couplers & (min, max) microseconds\\
            \hline
            \hline
            \texttt{DW\_2000Q\_6} & Chimera $C_{16}$ & 2041 & 5974 & (1, 2000)\\
            \hline
            \texttt{Advantage\_system4.1} & Pegasus $P_{16}$ & 5627 & 40279 & (0.5, 2000)\\
            \hline
            \texttt{Advantage\_system6.1} & Pegasus $P_{16}$ & 5616 & 40135 & (0.5, 2000)\\
            \hline
            \texttt{Advantage2\_prototype1.1} & Zephyr $Z_{4}$ & 563 & 4790 & (1, 2000)\\
            \hline
        \end{tabular}
    \end{center}
    \caption{D-Wave Quantum Annealing processor summary. }
    \label{tab:hardware_summary}
\end{table}

The posiform planting method requires solving a 2-SAT problem repeatedly during the planting process in order to verify the uniqueness of the planted solution, see Section~\ref{sec:generation}. For efficiency reasons, we add an initial batch of $B$ clauses to the 2-SAT problem before starting to check for uniqueness. In Section~\ref{sec:results_native_QUBO}, we employ the choice $B=2000$ for the Chimera hardware graph of \texttt{DW\_2000Q\_6}, $B=30000$ for the Pegasus hardware graph of \texttt{Advantage\_system4.1} and \texttt{Advantage\_system6.1}, and $B=1000$ for the Zephyr hardware graph of \texttt{Advantage2\_prototype1.1}. In Section~\ref{sec:results_minor_embedded}, we employ $B=1$ to generate the $52$ variable all-to-all graphs. These choices of $B$ are arbitrary, they do not influence the uniqueness of the solution but the runtime of the generation process, and they were selected to correspond to the number of variables in the hardware graph. Likewise, the posiform coefficients can be chosen arbitrarily in posiform planting. We select the posiform coefficients from the set $\{1, 2\}$ for both the hardware native QUBOs and the minor embedded QUBOs, which depending on the hardware graph can result in highly variable QUBO coefficients after converting the posiform to a QUBO. However, the QUBO models can still be mapped onto the D-Wave hardware due to the auto coefficient scaling and the maximum energy scale that is programmable onto the chip. Choosing the posiform coefficients as integers also ensures that the QUBO coefficients will be integers. Visualizations of the hardware native QUBO coefficients can be found in Appendix~\ref{sec:appendix_QUBO_hardware}.

The hardware native QUBOs in Section~\ref{sec:results_native_QUBO} are sampled using annealing times of $0.5$ microseconds for the \texttt{Advantage\_system6.1} and \texttt{Advantage\_system4.1}, and in the range $\{1, 2, \dots, 10\}$ as well as $\{20, 30, \dots, 1990, 2000\}$ microseconds for all four D-Wave quantum annealers. Each hardware native QUBO is sampled using two D-Wave device calls, each having $400$ anneal-readout cycles, resulting in a total of $800$ measurements made per annealing time and per hardware native QUBO.

\subsection{Results for hardware native QUBOs}
\label{sec:results_native_QUBO}
We generate $100$ unique QUBO problems tailored to the four D-Wave quantum annealers outlined in Table~\ref{tab:hardware_summary}. Those are being solved as a function of the anneal time, using the D-Wave settings described in Section~\ref{sec:results_parameter_settings}. Since the unique solution and thus the ground state of each QUBO is known, computing the ground state success probability (GSP) is straightforward.

\begin{figure}[h]
    \centering
    \includegraphics[width=0.49\textwidth]{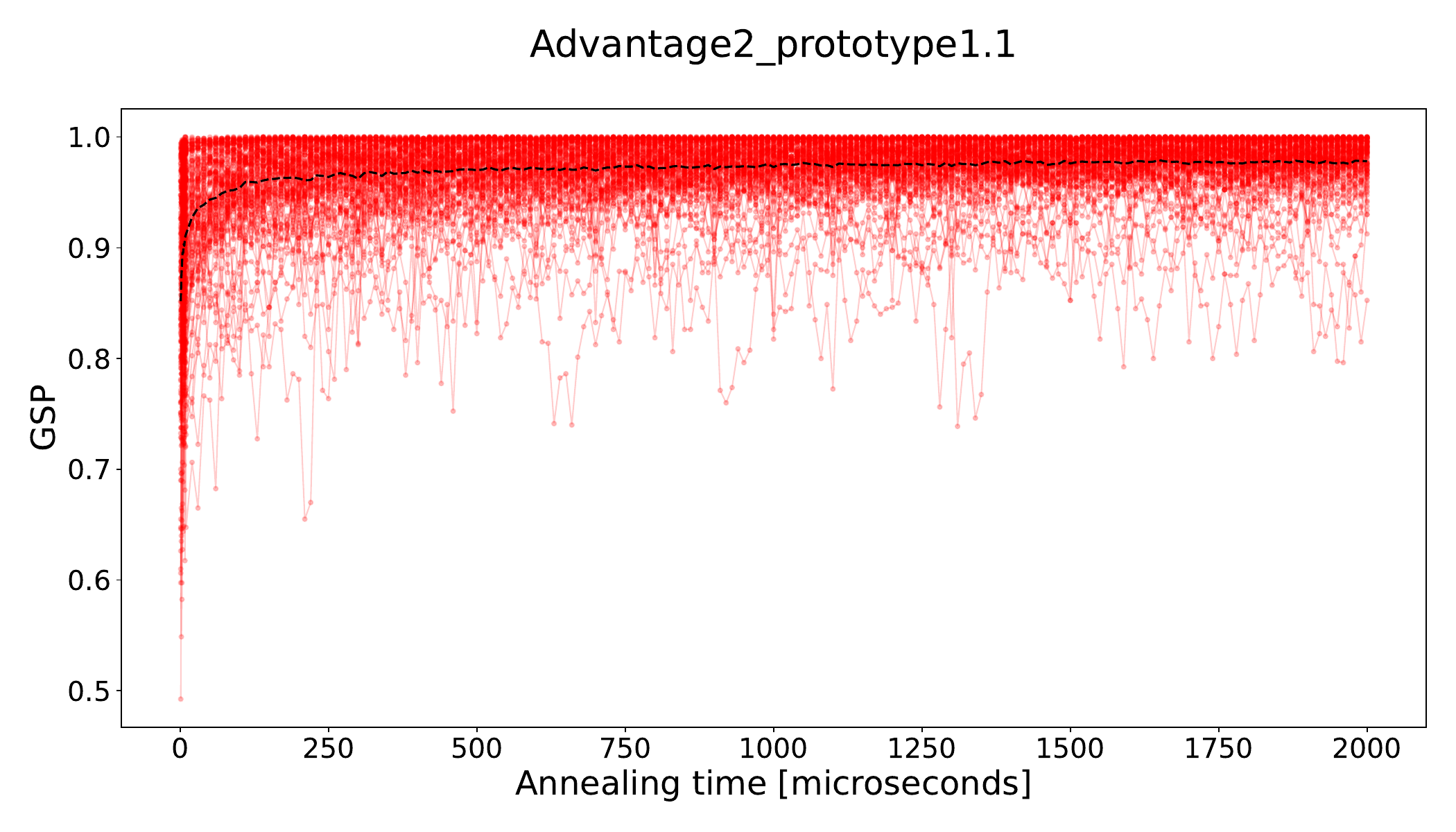}
    \includegraphics[width=0.49\textwidth]{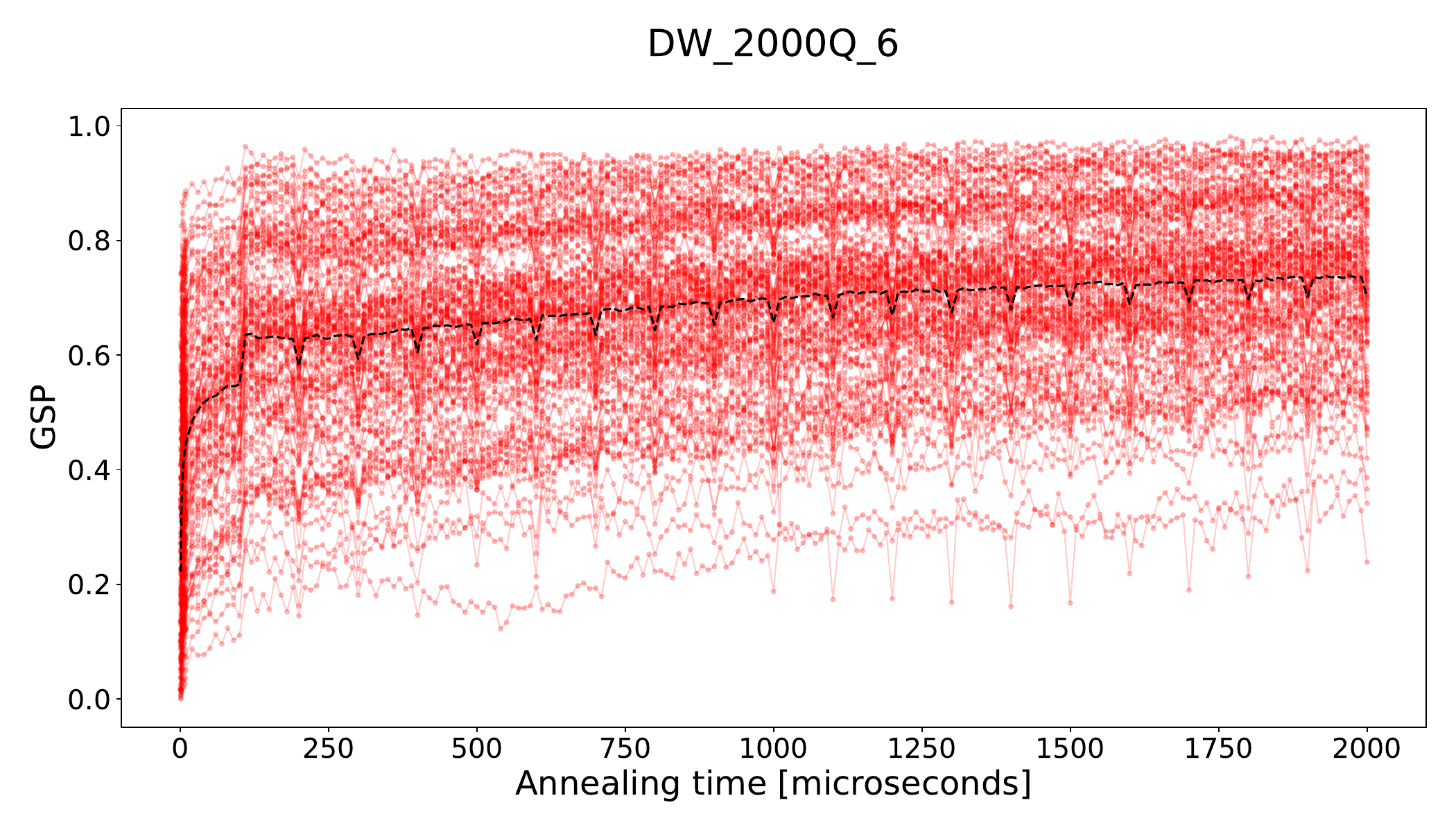}\\
    \includegraphics[width=0.49\textwidth]{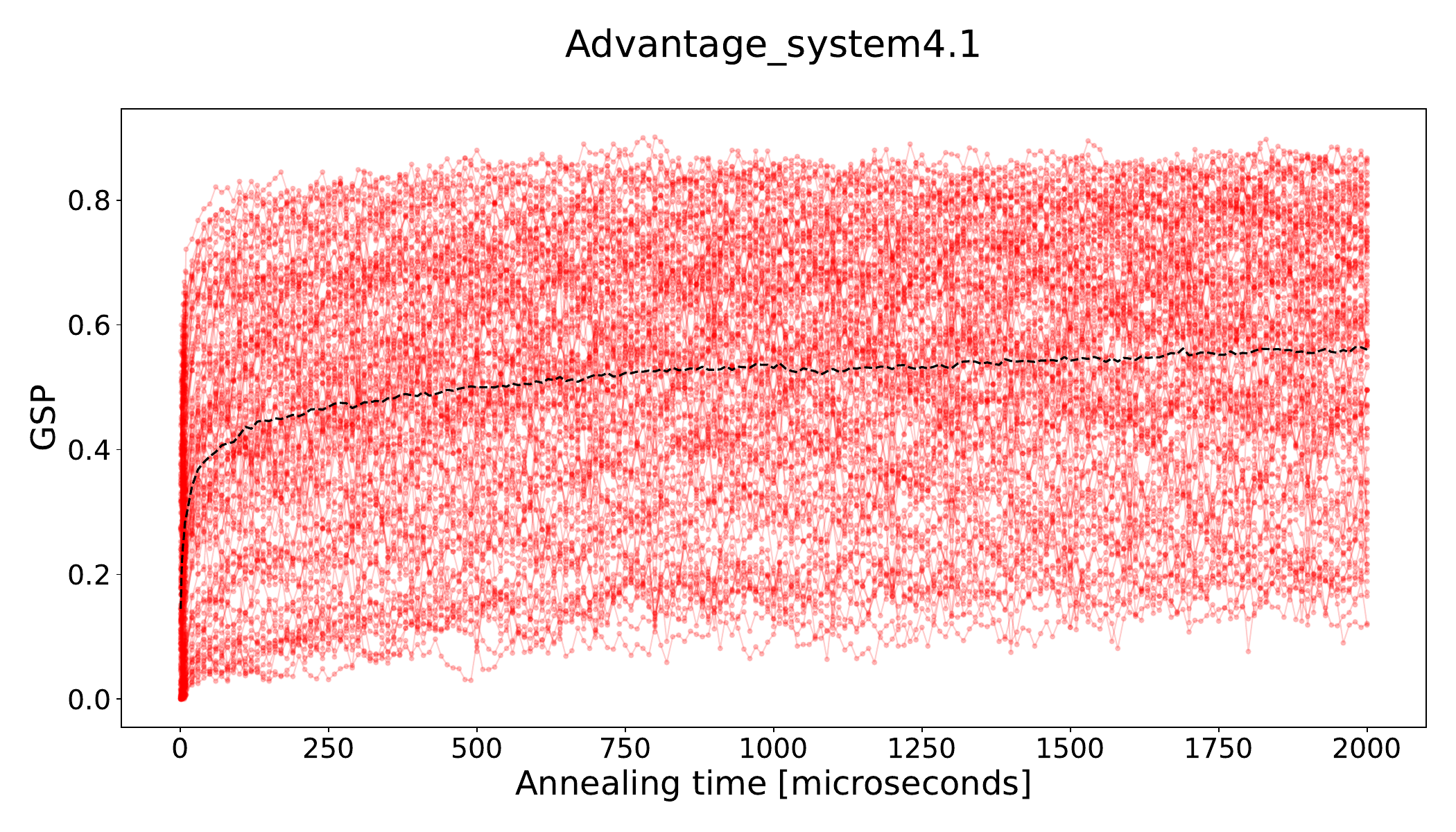}
    \includegraphics[width=0.49\textwidth]{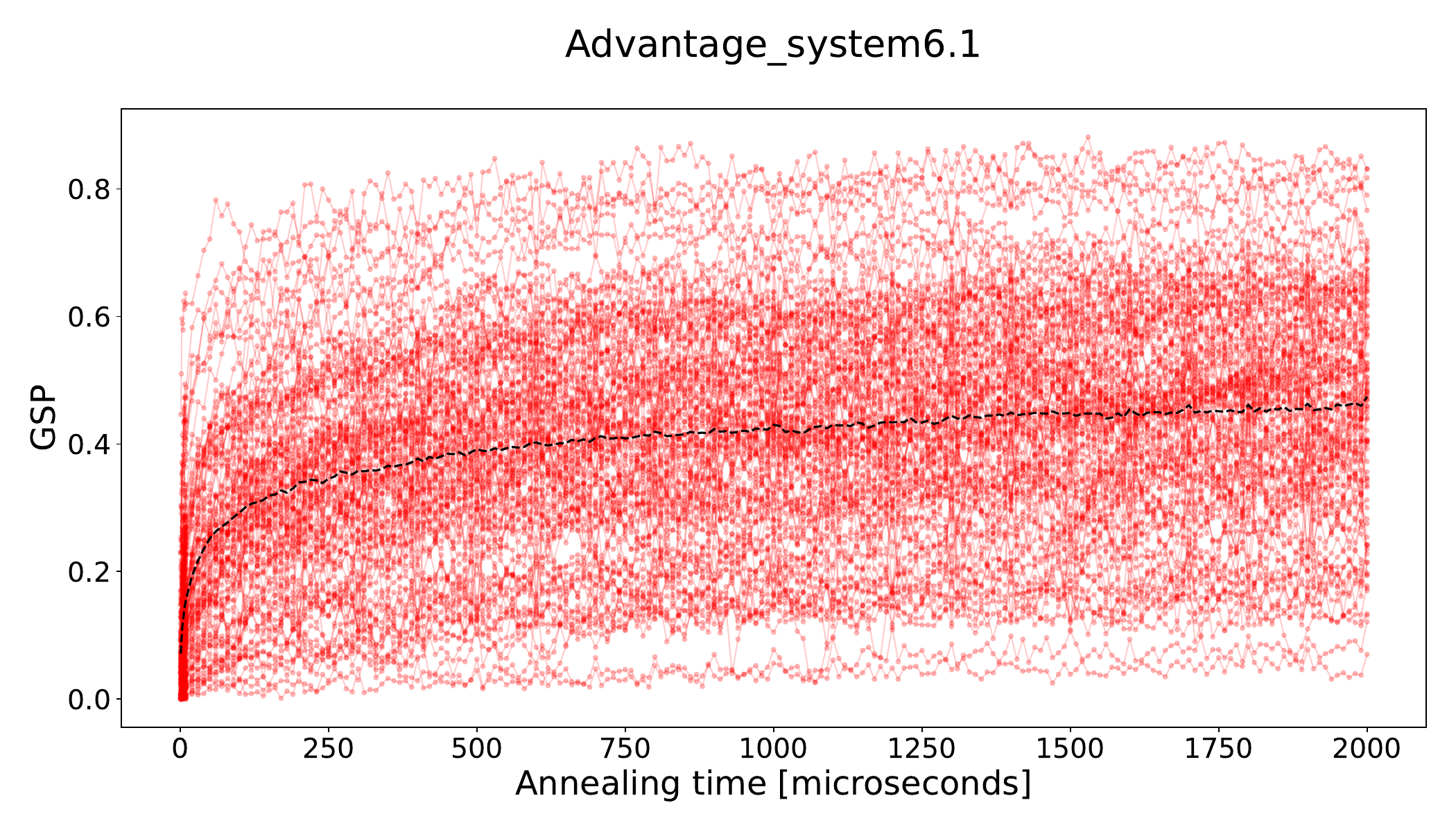}
    \caption{Ground state success probability (GSP) for hardware native QUBOs computed on the devices \texttt{Advantage2\_prototype1.1} (top left), \texttt{DW\_2000Q\_6} (top right), \texttt{Advantage\_system4.1} (bottom left), and \texttt{Advantage\_system6.1} (bottom right). Each subplot corresponds to one D-Wave annealer and contains $100$ separate lines which are showing GSP results for the $100$ unique random hardware native posiform planted QUBOs. Each line shows the probability of reaching the ground state (among the $800$ anneals) as a function of the annealing time. The dashed black line denotes the mean GSP computed at each evaluated annealing time.}
    \label{fig:native_QUBO_GSP}
\end{figure}

Figure~\ref{fig:native_QUBO_GSP} shows the GSP for the hardware native QUBOs measured on the four D-Wave devices. Each subplot shows the results of the $100$ randomly generated QUBOs on each device, with one line per QUBO visualizing probability of reaching the ground state (among the $800$ anneals) as a function of the annealing time.

Several observations are noteworthy. Since the GSP is mostly non-zero, the D-Wave quantum annealers are able to sample the optimal solution during some anneal. This even holds true for QUBO instances with up to $5627$ variables in the case of \texttt{Advantage\_system4.1}). Although it is difficult to see in the plots, at small annealing times, in particular $500$ nanoseconds and $1$ microsecond, the two Pegasus chip devices fail to sample the optimal solution across all $100$ problem instances.

We observe an increasing trend in the measured GSP as a function of the annealing time, but with diminishing returns as annealing time increases. The results show a difference in behavior between the four D-Wave devices. In particular, the $563$ qubit system \texttt{Advantage2\_prototype1.1} samples the optimal solution at a much higher rate than the other devices. This finding can be attributed to the fact that the number of variables on this device is less than on the other devices, while also being the newest generation of the D-Wave annealer with reported lower error rates than the previous generations.

We observe that the results for \texttt{DW\_2000Q\_6} in Figure~\ref{fig:native_QUBO_GSP} show periodic variations of the measured GSP. This is because the annealing time measurements in increments of $100$ microseconds were made several weeks apart from the measurements made for all other annealing times in increments of $10$ microseconds, and previous studies \cite{Pelofske_2023} have shown that there are long term variations (in solution quality) of the computations carried out on current D-Wave quantum annealing devices. Therefore, the variations that have a periodicity of $100$ microseconds are due to variance of the noise profile of the device, rather than variations that are a function of the annealing time. 

Next we examine the \textit{time-to-solution} (\textit{TTS}) metric for the $100$ QUBO instances that were generated for each of the four D-Wave annealers. TTS is an estimate of the time it takes to reach an optimum solution with a 99 percent confidence. It is defined as
\begin{align}
    \text{TTS}_{0.99} = \frac{\text{QPU-access-time}}{A} \cdot  \frac{\log(1-0.99)}{\log(1-p)},
    \label{eq:TTS}
\end{align}
where QPU-access-time (in seconds) is the real compute time used on the D-Wave backend (including the hardware programming time, anneal-readout cycle, and anneal times), $A$ is the number of anneals, and $p \in (0,1)$ is the success probability observed among the $A$ anneals, that is, the proportion of anneals that found the ground state. The QPU-access-time also includes all communication time with the device on top of the annealing time used in the computation. When $p=1$, we set $\text{TTS}_{0.99} = \text{QPU-access-time}/A$. When $p=0$, $\text{TTS}_{0.99}$ is undefined, and therefore is not computed.

\begin{figure}[h]
    \centering
    \includegraphics[width=0.49\textwidth]{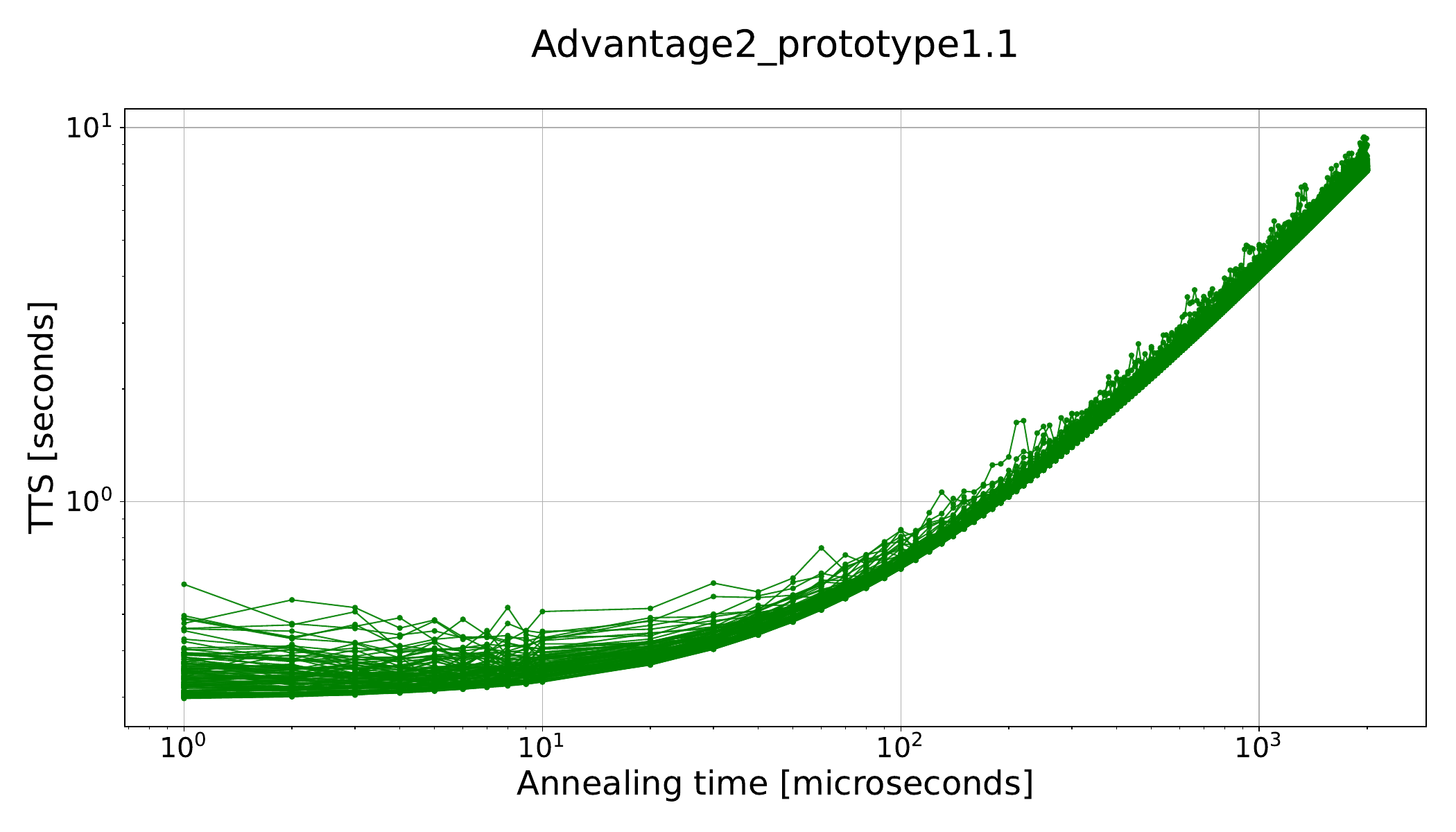}
    \includegraphics[width=0.49\textwidth]{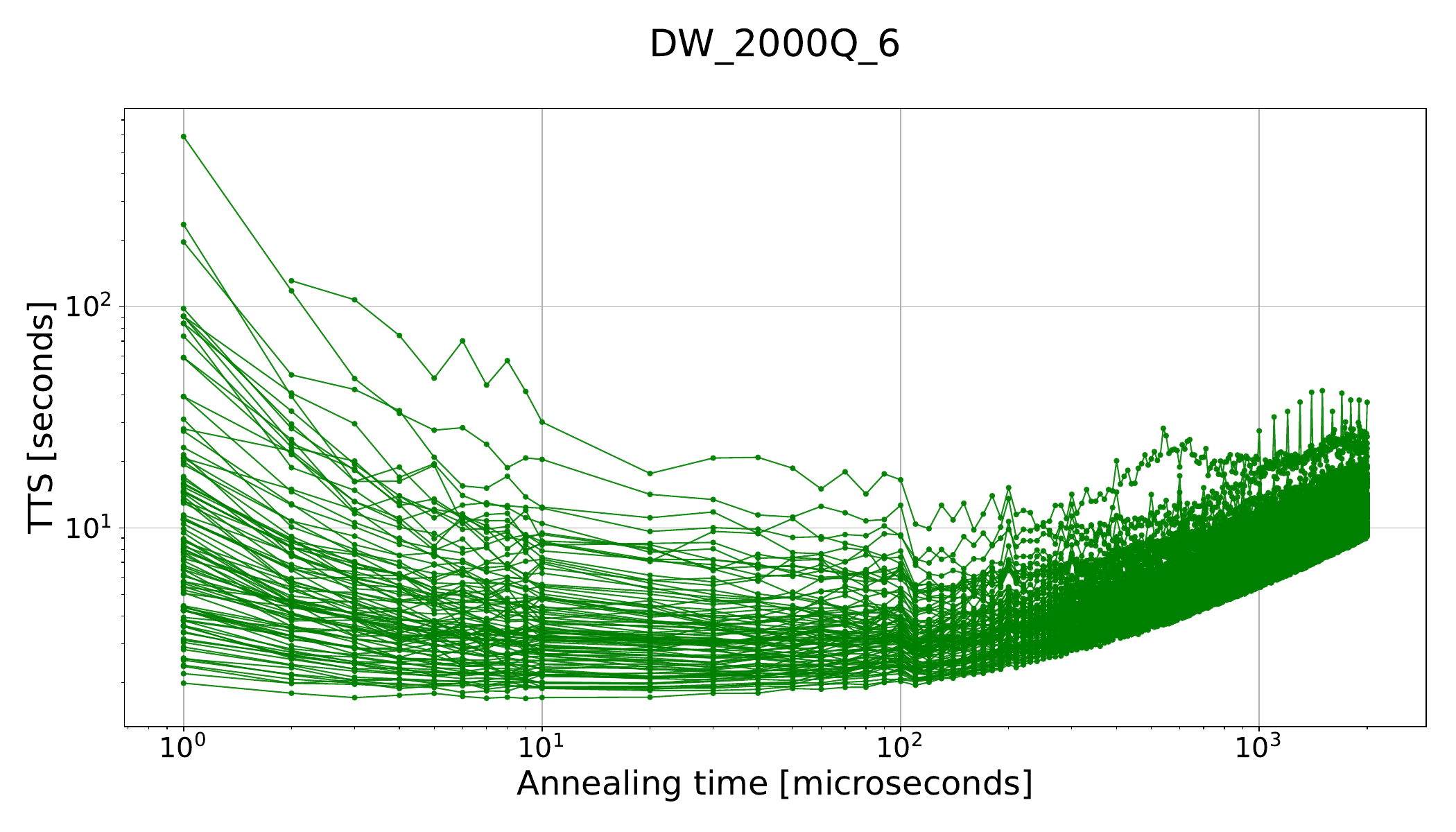}
    \includegraphics[width=0.49\textwidth]{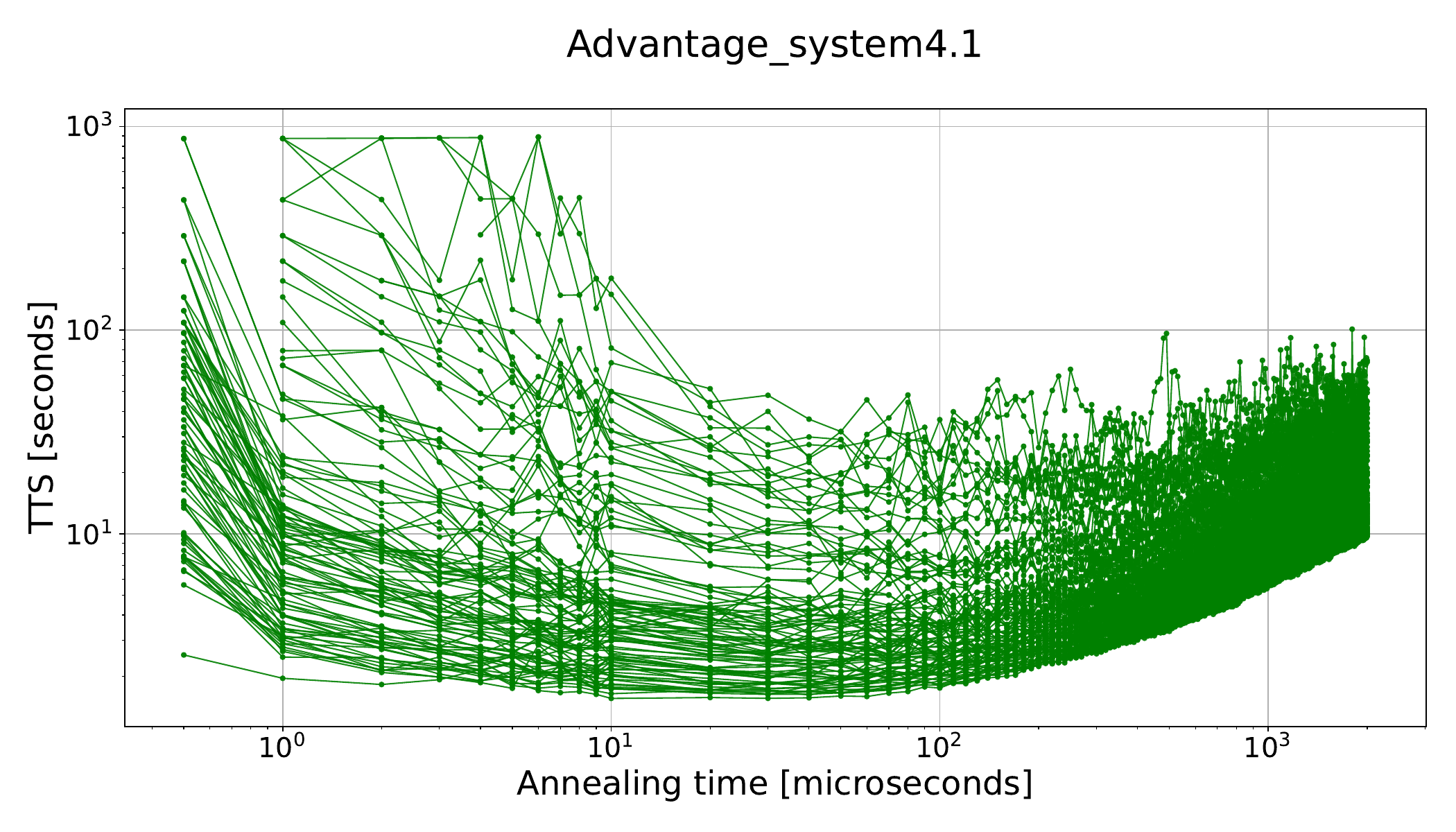}
    \includegraphics[width=0.49\textwidth]{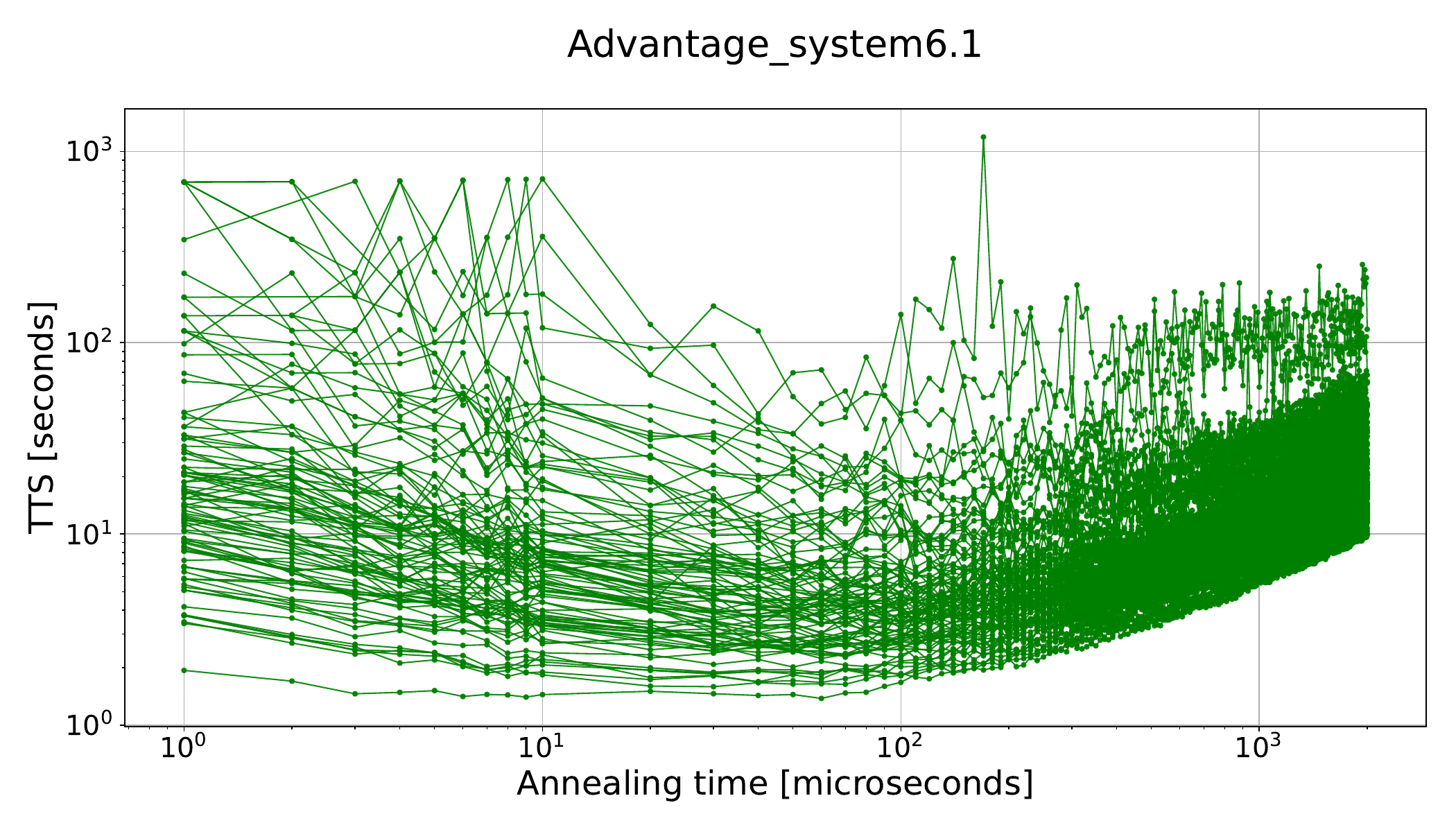}
    \caption{TTS as a function of the annealing time on the $100$ hardware QUBO problems for each of the four D-Wave quantum annealers, in particular \texttt{Advantage2\_prototype1.1} (top left), \texttt{DW\_2000Q\_6} (top right), \texttt{Advantage\_system4.1} (bottom left), and \texttt{Advantage\_system6.1} (bottom right). One line per QUBO instance. Log scale on both axes.}
    \label{fig:native_QUBO_TTS}
\end{figure}

Figure~\ref{fig:native_QUBO_TTS} plots the TTS (computed with eq.~\eqref{eq:TTS}) to reach the optimal planted solution for the set of $100$ randomly generated hardware native QUBOs for each of the four D-Wave annealers. We observe that for \texttt{Advantage2\_prototype1.1}, the lowest TTS is achieved for short annealing times, whereas for the other three generations of the D-Wave annealer, both low and high annealing times incur higher TTS values, with the lowest TTS being achieved in-between.

\begin{figure}[h]
    \centering
    \includegraphics[width=0.49\textwidth]{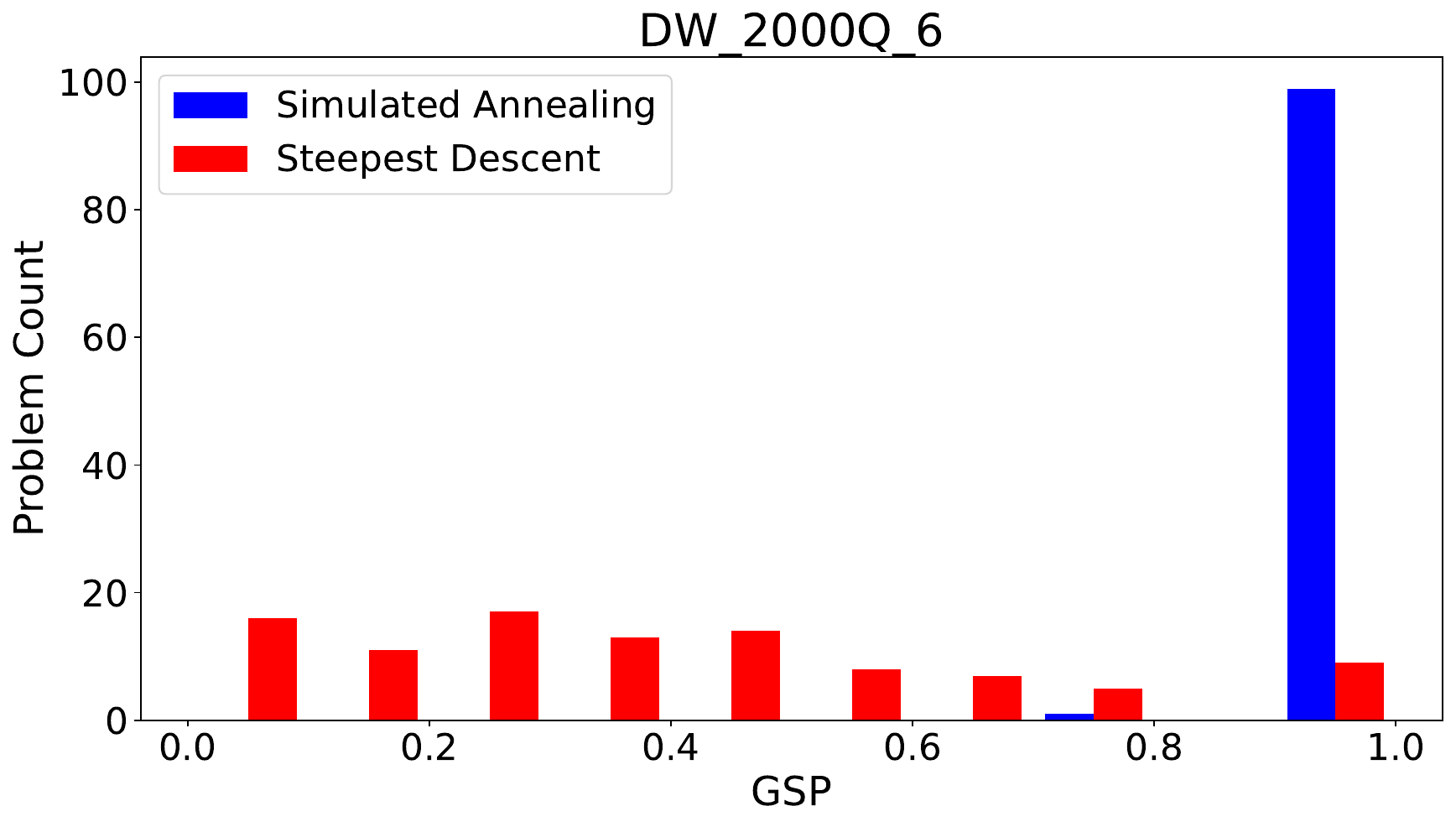}
    \includegraphics[width=0.49\textwidth]{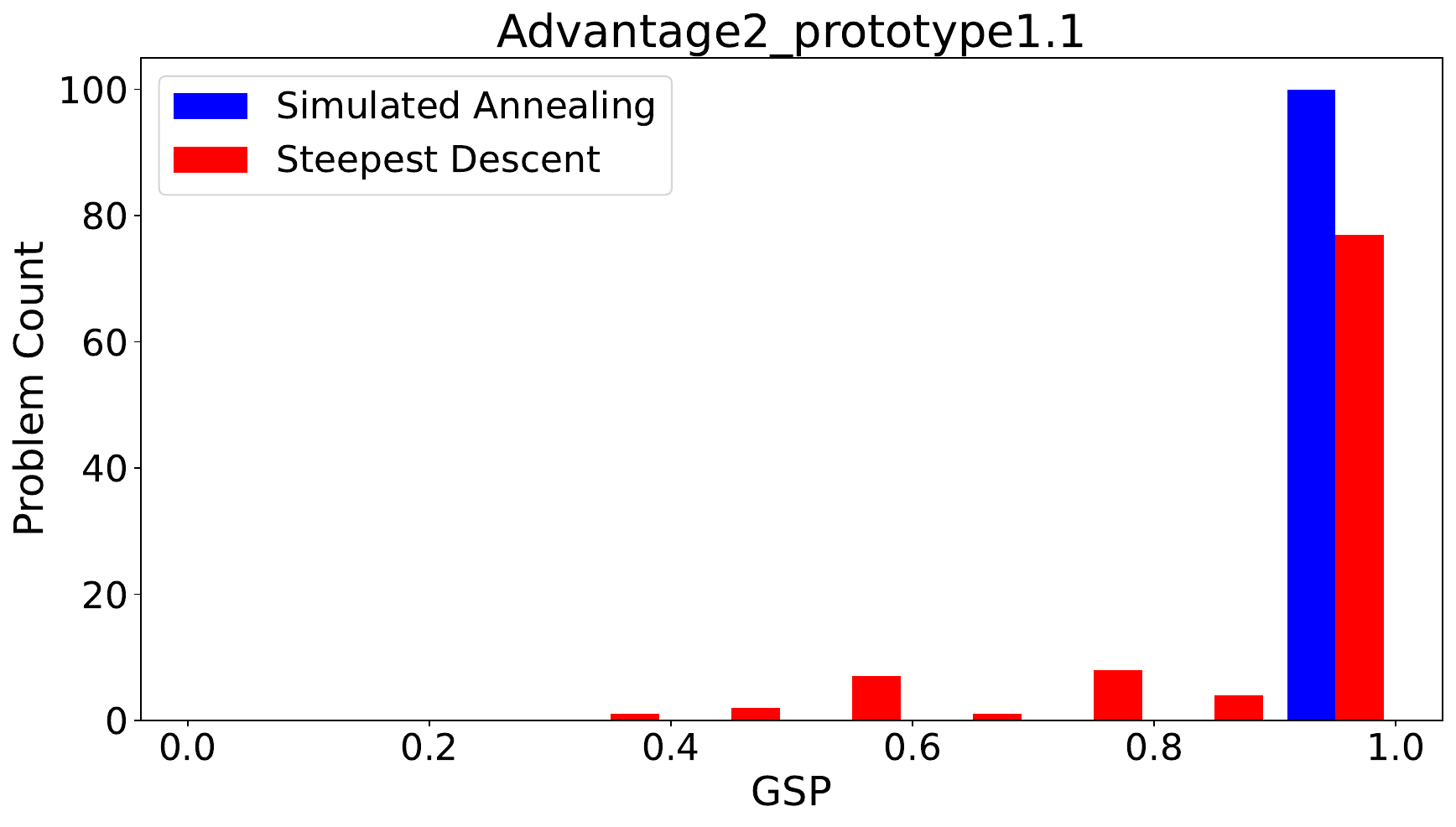}
    \includegraphics[width=0.49\textwidth]{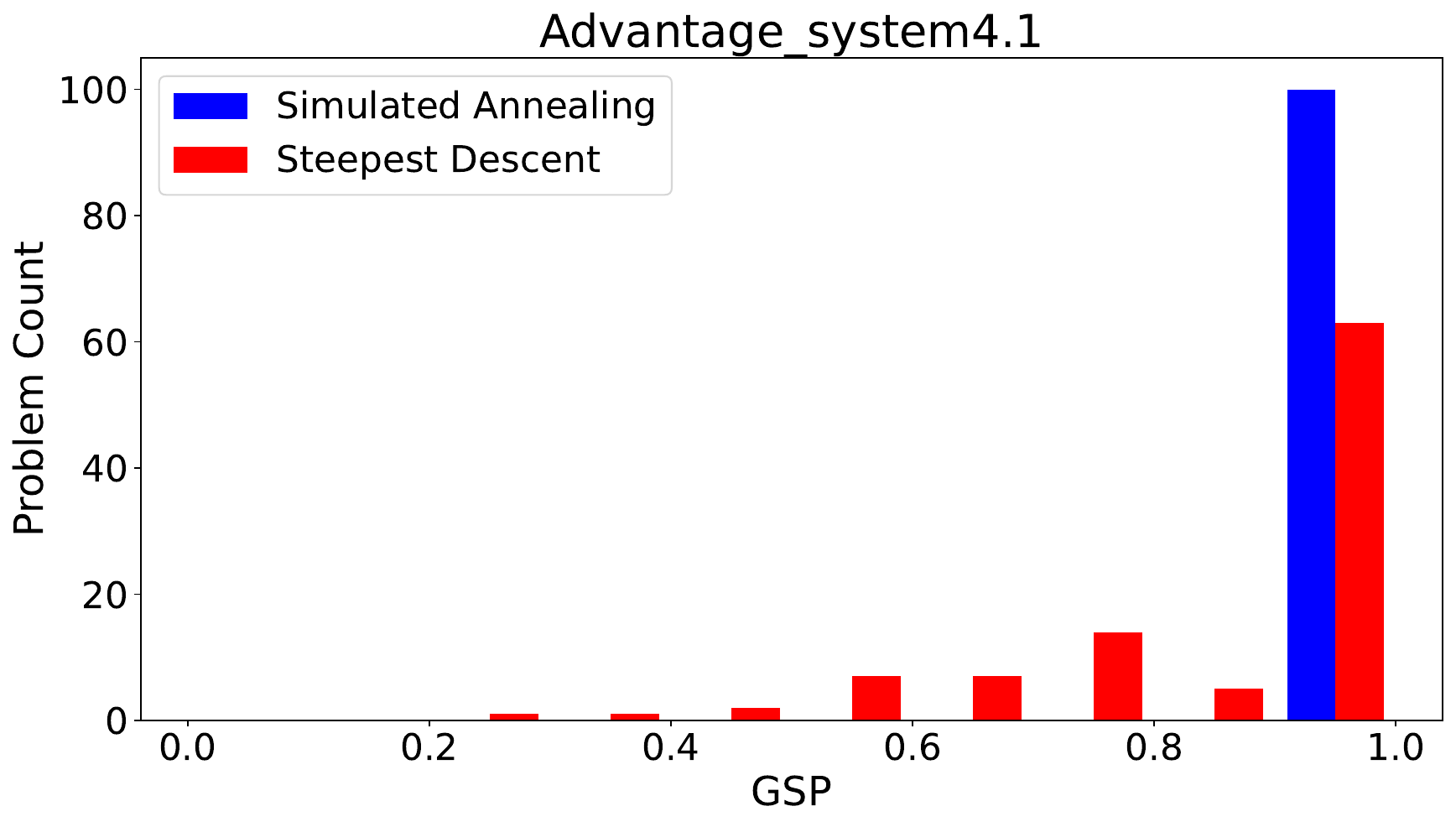}
    \includegraphics[width=0.49\textwidth]{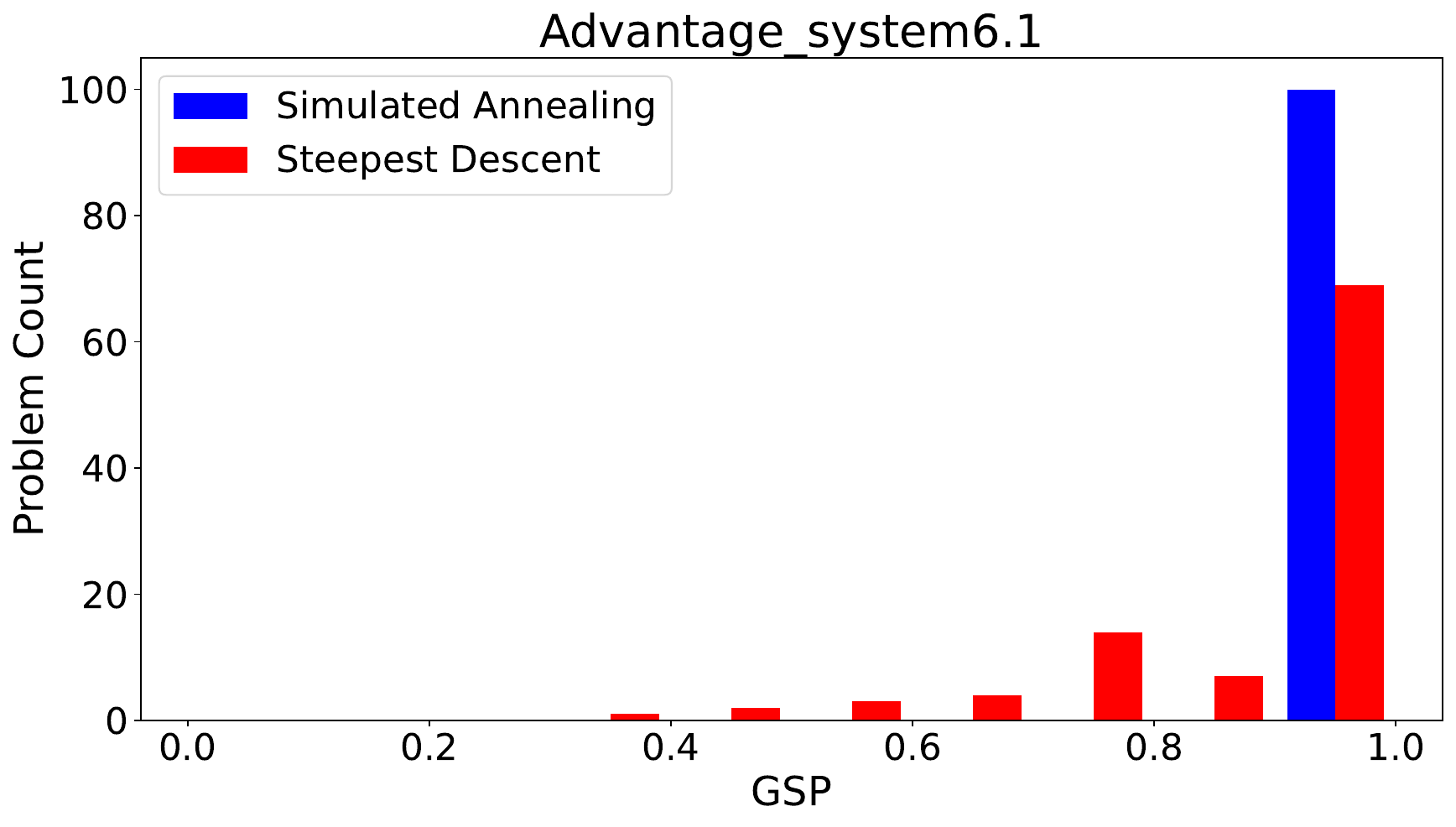}
    \caption{Histograms of the GSP for the same hardware native posiform planted QUBOs used in Figure~\ref{fig:native_QUBO_GSP} sampled using classical heuristics. Simulated annealing and steepest descent heuristics applied to the QUBOs generated for the hardware graphs of \texttt{DW\_2000Q\_6} (top left), \texttt{Advantage2\_prototype1.1} (top right), \texttt{Advantage\_system4.1} (bottom left), \texttt{Advantage\_system6.1} (bottom right). The side-by-side histogram bars correspond to each bin, so the sampling rates for simulated annealing are extremely high (usually at a proportion of $1$). }
    \label{fig:native_QUBO_GSP_classical_heuristics}
\end{figure}

In addition to solving the $100$ native hardware QUBOs sampled on each of the four D-Wave devices, we also investigate how successfully classical heuristics can solve them. Figure~\ref{fig:native_QUBO_GSP_classical_heuristics} shows histograms for the achieved GSP when sampling the same set of hardware native QUBO problems using the classical heuristics Simulated Annealing (SA), implemented in the function \textit{neal} in the D-Wave SDK \cite{neal} and greedy Steepest Descent (analogous to steepest gradient descent), implemented in the function \textit{greedy} in the D-Wave SDK \cite{greedy}. Figure~\ref{fig:native_QUBO_GSP_classical_heuristics} demonstrates that the SA algorithm in particular is able to find the optimal solution of the QUBOs generated with posiform planting with very high success probability.

\subsection{Results for minor embedded QUBOs}
\label{sec:results_minor_embedded}
Posiform planting as introduced in Section~\ref{sec:generation} generates new clauses to be added to the 2-SAT instance without any constraints on the indices. Though clauses are arbitrary, the generated QUBOs are usually not fully connected. Using generated QUBOs with all-to-all connectivity require a minor embedding onto the D-Wave quantum hardware before being solved, since the D-Wave hardware graphs are (relatively) sparse. Despite the challenges associated with minor-embedded QUBO instances, which require chained qubits and pose issues such as selecting appropriate chain strengths, utilizing such QUBOs enables a direct comparison of D-Wave devices on the same set of input problems. A diagram showing these complete minor embeddings on the four hardware graphs is shown in Figure~\ref{fig:52_minor_embedding} in the appendix.

We generate $5$ QUBOs with varying density with the aim to allow for a range of GSP rates among those planted QUBOs. Each QUBO instance has $52$ logical variables, which is the largest problem size with an all-to-all connectivity that can be minor embedded on the \texttt{Advantage2\_prototype1.1} device. Since \texttt{Advantage2\_prototype1.1} has the smallest such embedding, the same QUBO instances are guaranteed to be executed on all four D-Wave quantum annealers, thereby allowing for a fair comparison. Note that these $52$ variable QUBOs are not fully connected, but they are arbitrarily connected in that the generator can select arbitrary edges to include. 

\begin{figure}[h]
    \centering
    \includegraphics[width=0.19\textwidth]{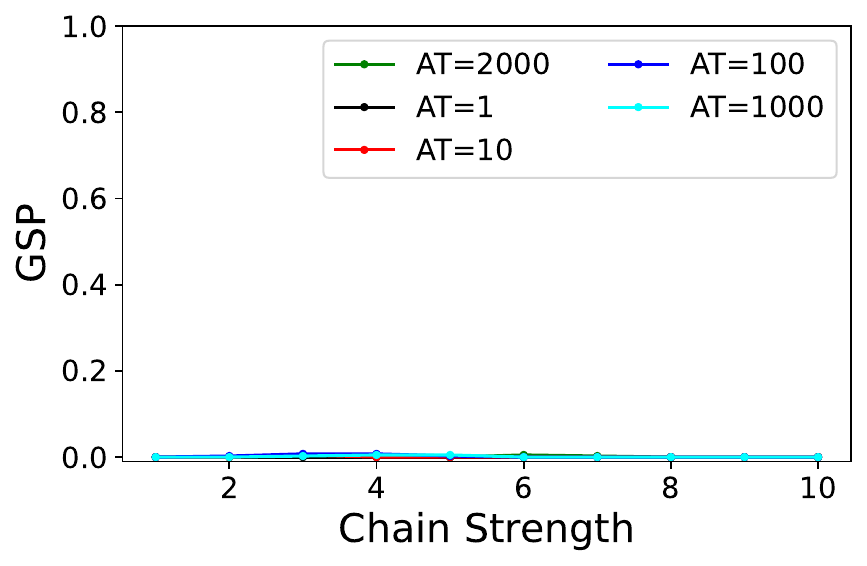}
    \includegraphics[width=0.19\textwidth]{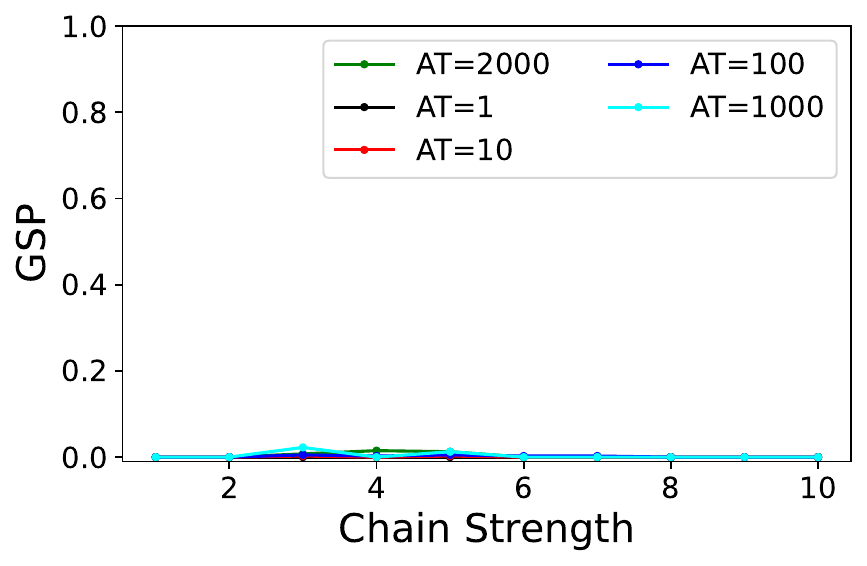}
    \includegraphics[width=0.19\textwidth]{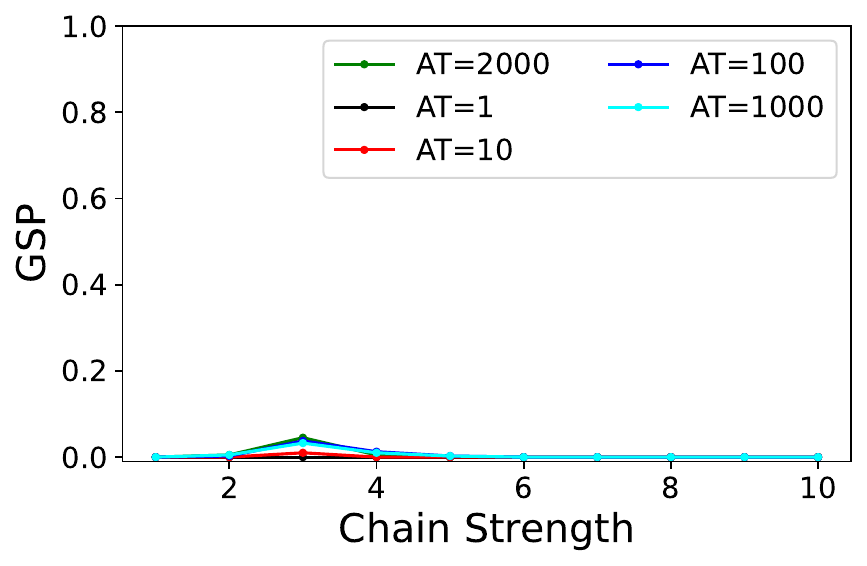}
    \includegraphics[width=0.19\textwidth]{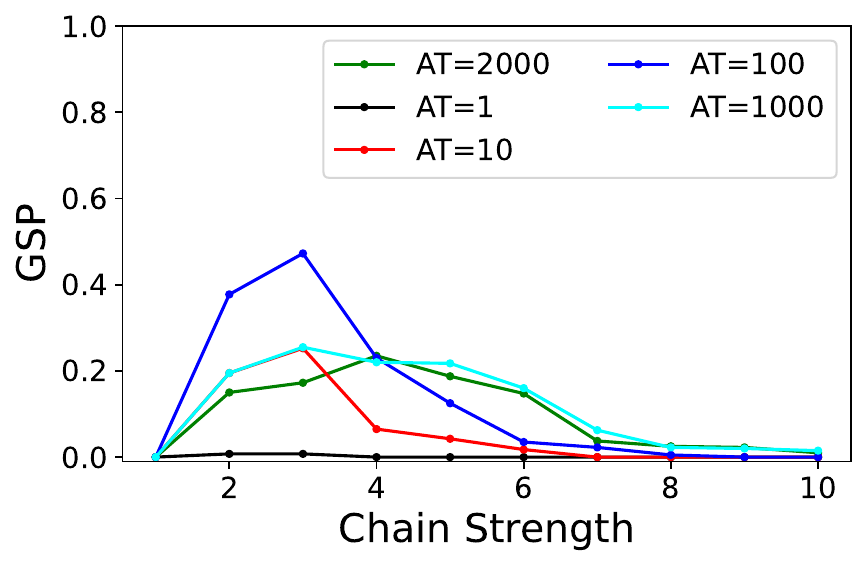}
    \includegraphics[width=0.19\textwidth]{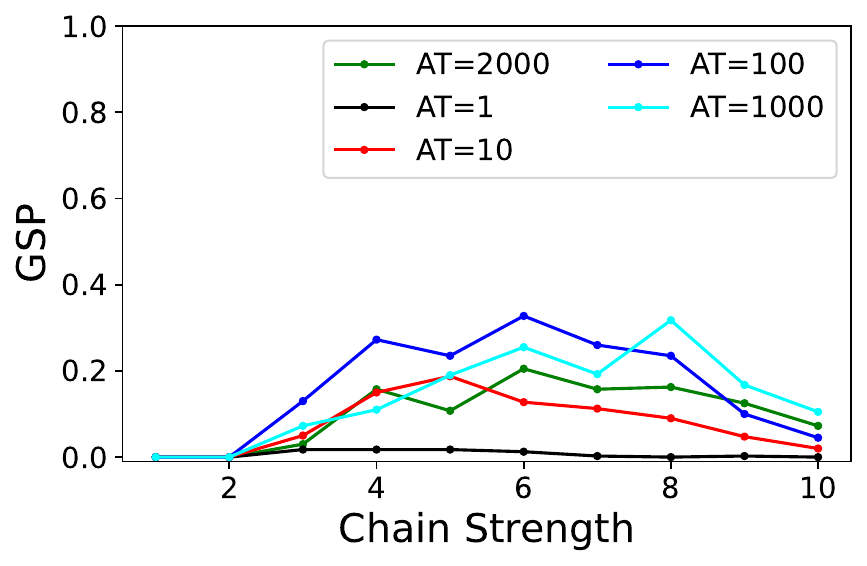}
    \includegraphics[width=0.19\textwidth]{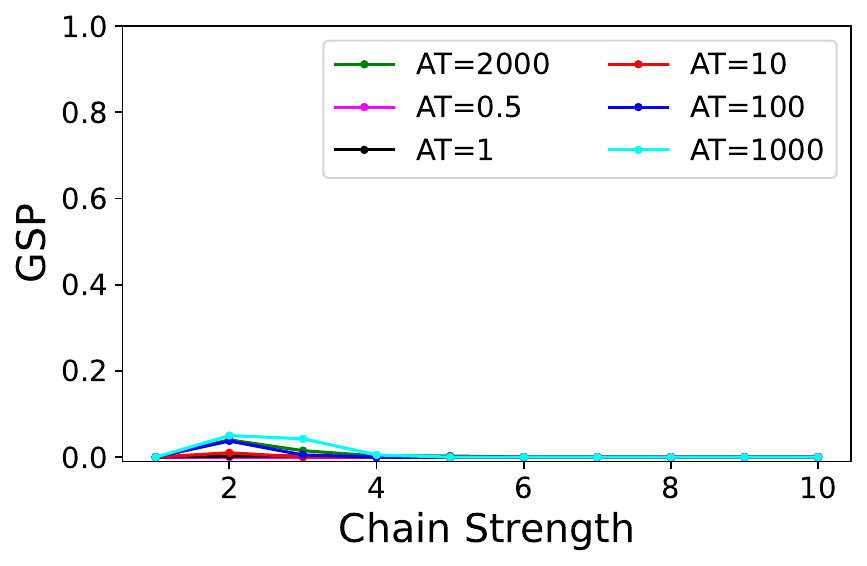}
    \includegraphics[width=0.19\textwidth]{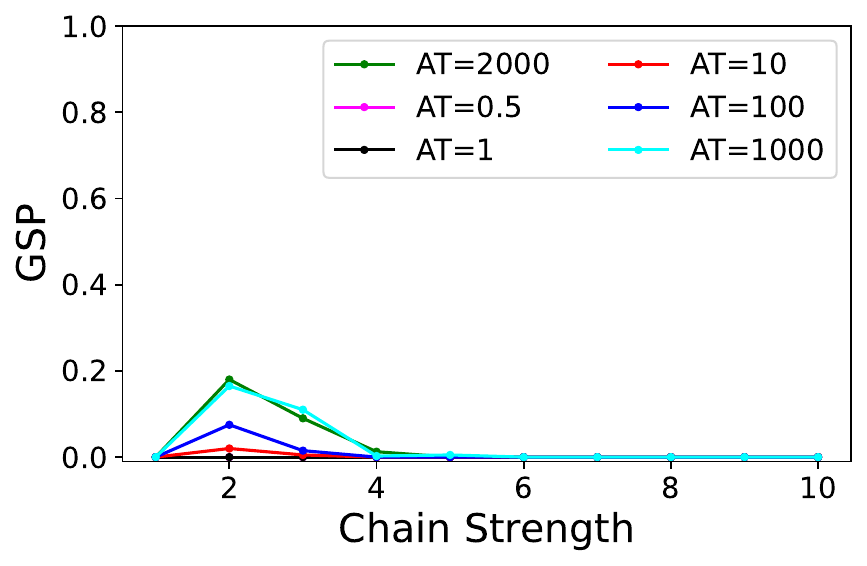}
    \includegraphics[width=0.19\textwidth]{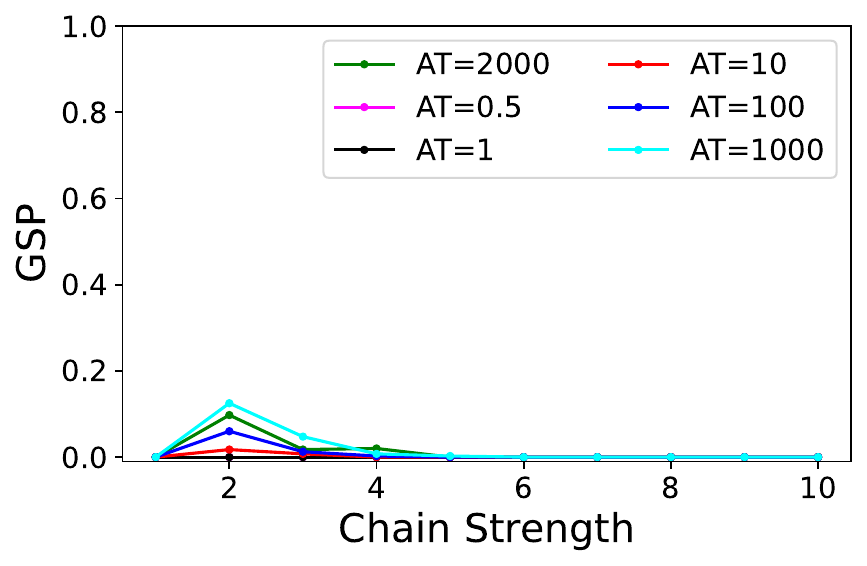}
    \includegraphics[width=0.19\textwidth]{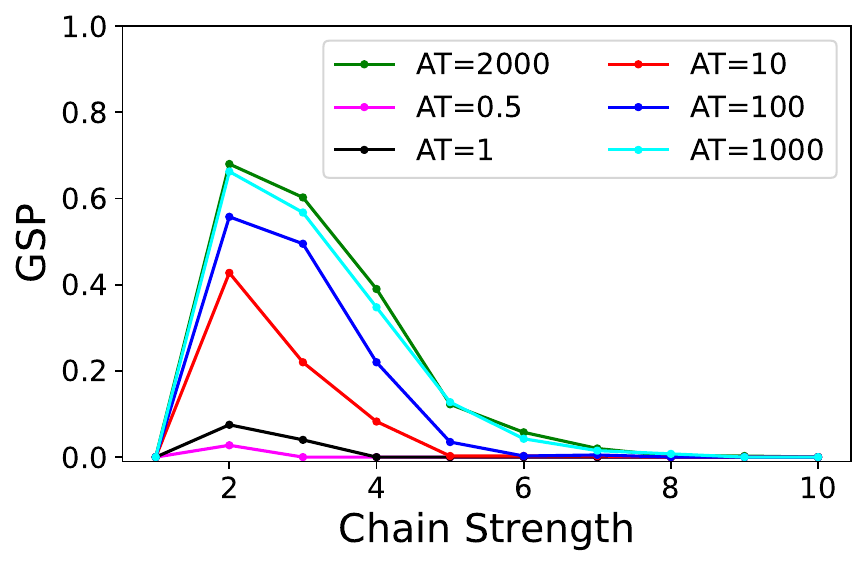}
    \includegraphics[width=0.19\textwidth]{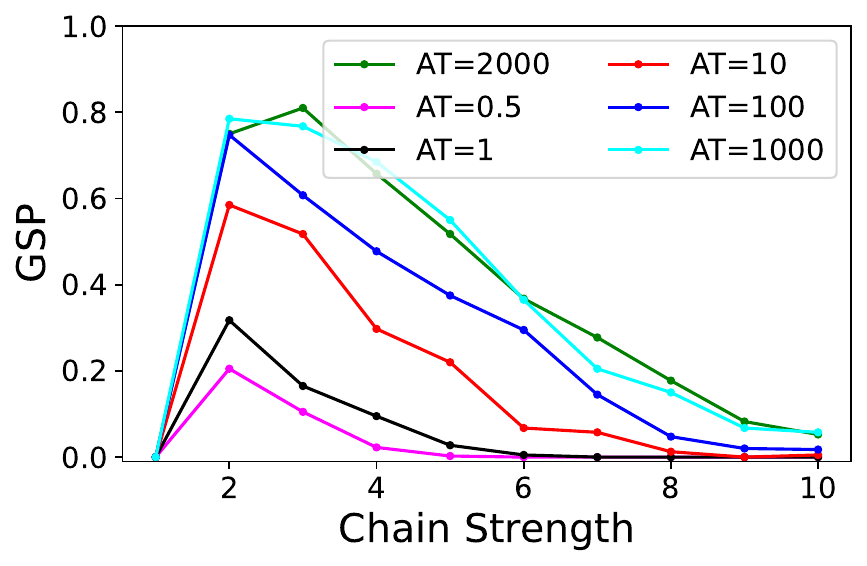}
    \includegraphics[width=0.19\textwidth]{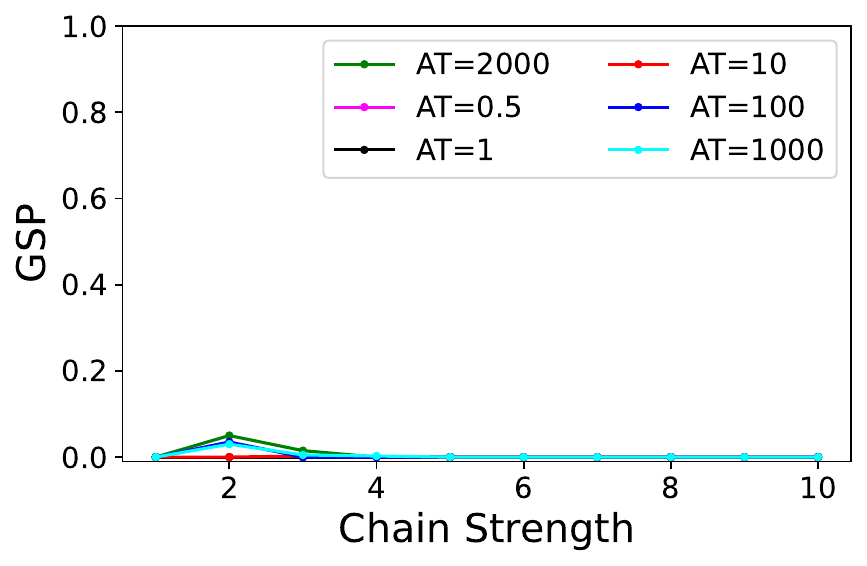}
    \includegraphics[width=0.19\textwidth]{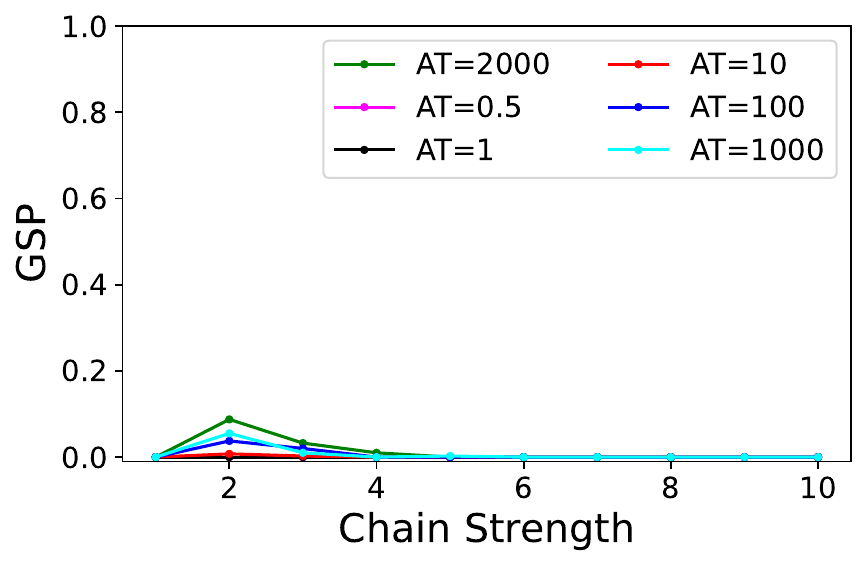}
    \includegraphics[width=0.19\textwidth]{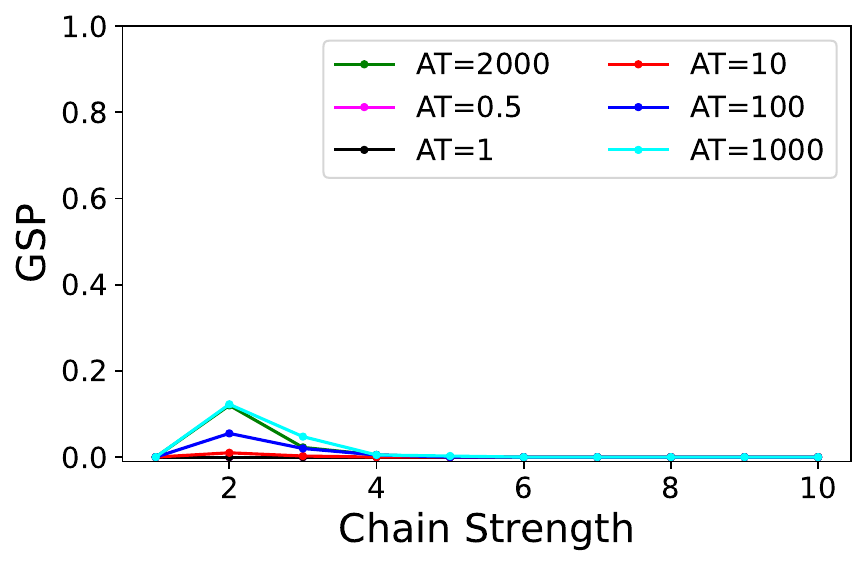}
    \includegraphics[width=0.19\textwidth]{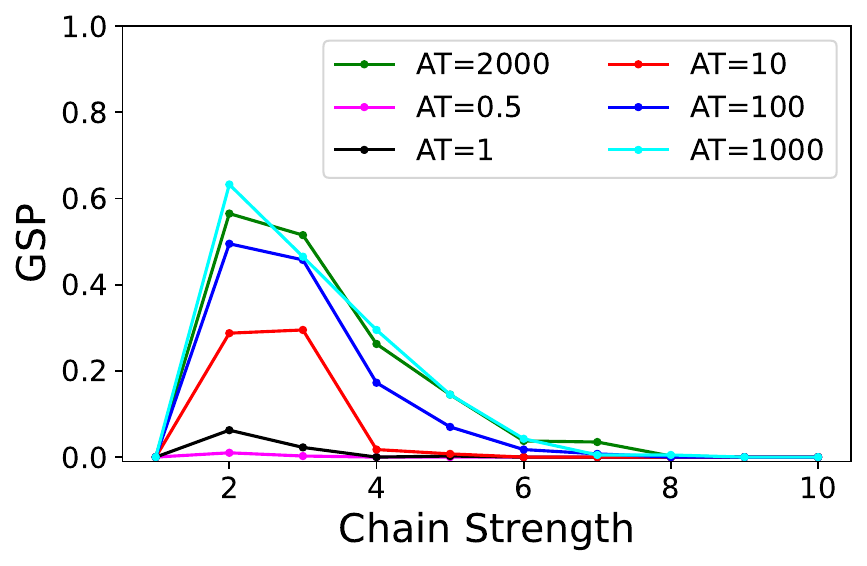}
    \includegraphics[width=0.19\textwidth]{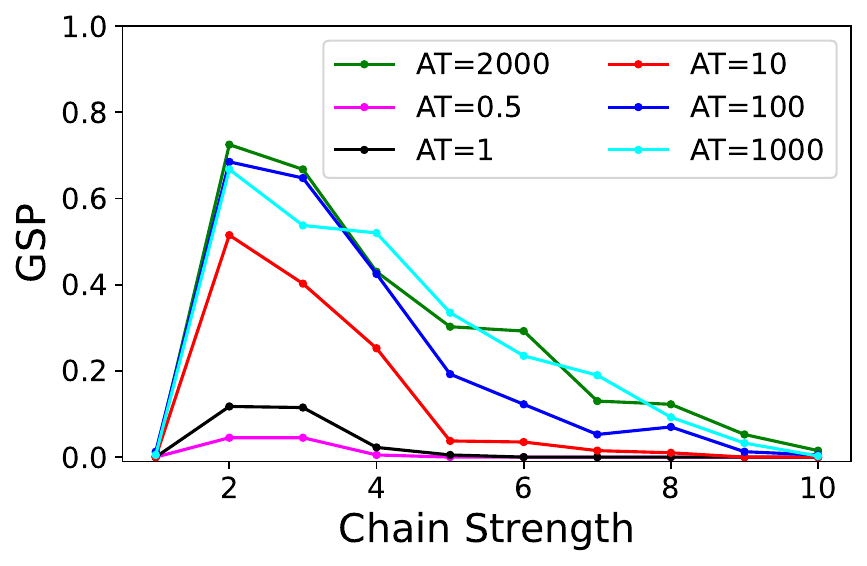}
    \includegraphics[width=0.19\textwidth]{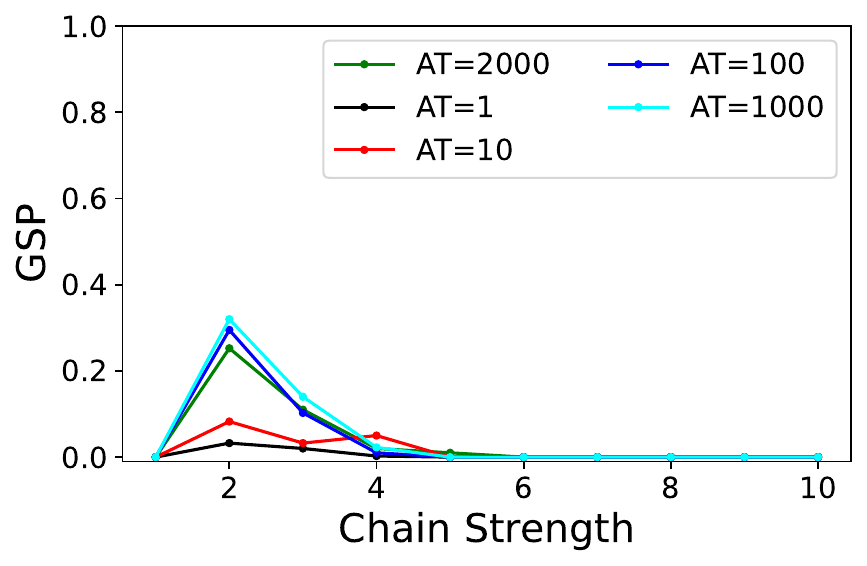}
    \includegraphics[width=0.19\textwidth]{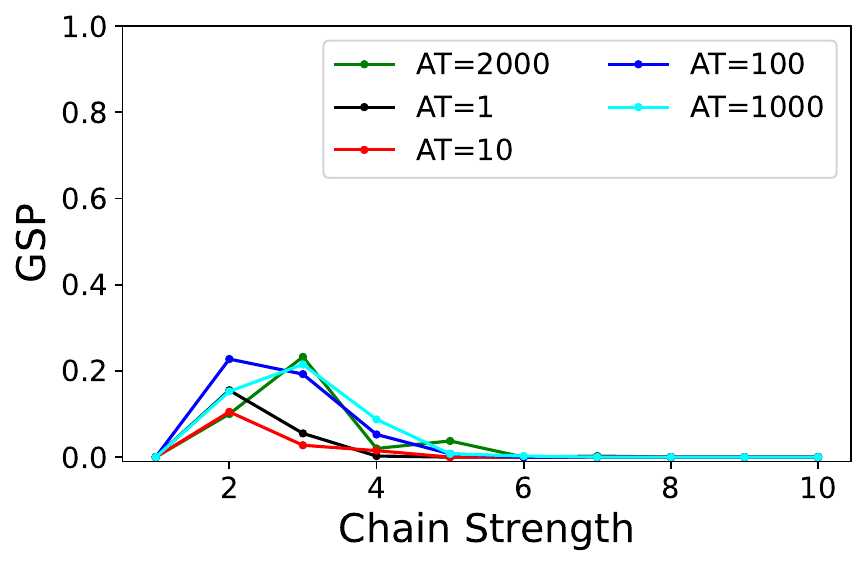}
    \includegraphics[width=0.19\textwidth]{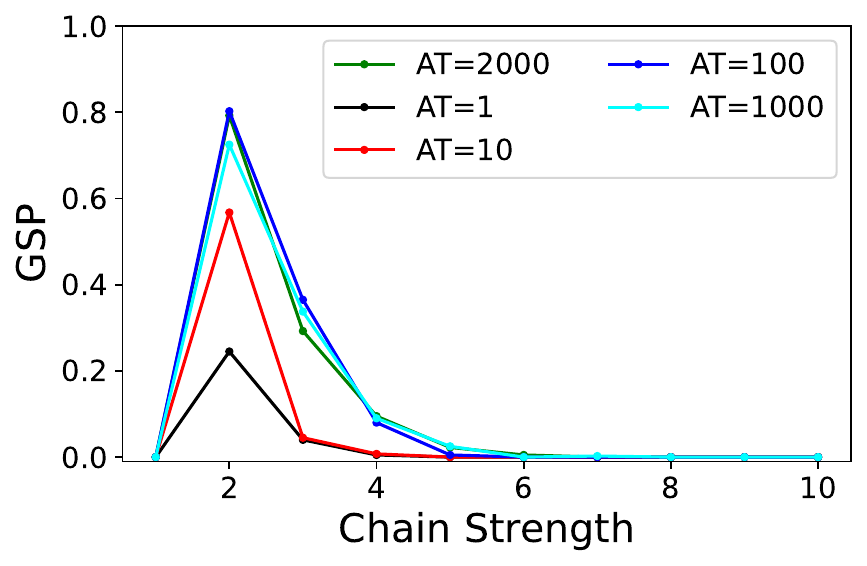}
    \includegraphics[width=0.19\textwidth]{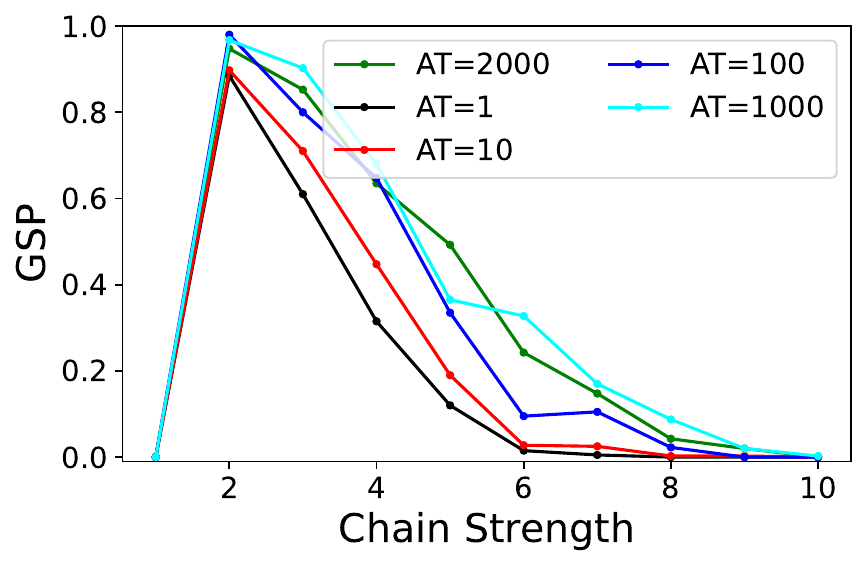}
    \includegraphics[width=0.19\textwidth]{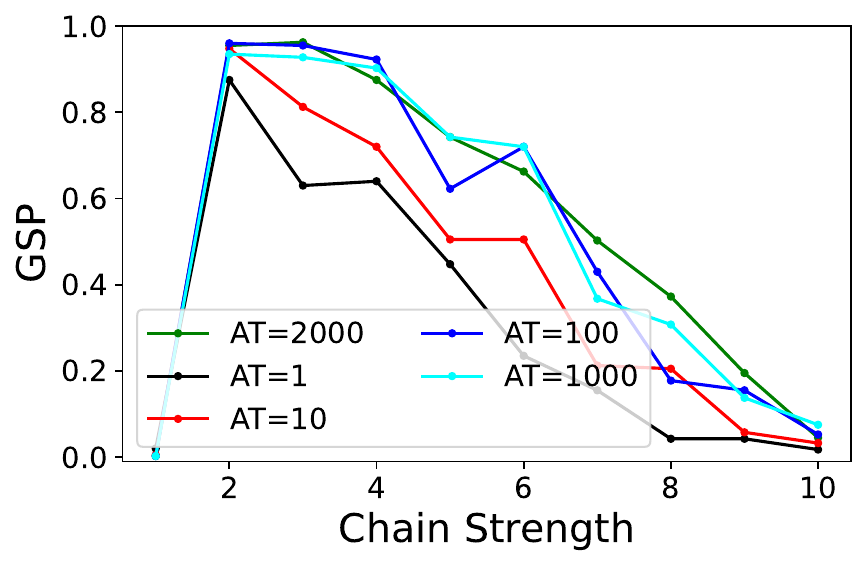}
    \caption{GSP measurements as a function of the chain strength for the $5$ minor-embedded QUBO instances. The $5$ columns correspond to the $5$ QUBO instances being solved, and the $4$ rows correspond to the $4$ D-Wave quantum annealers (\texttt{DW\_2000Q\_6}, \texttt{Advantage\_system4.1}, \texttt{Advantage\_system6.1}, \texttt{Advantage2\_prototype1.1} from top to bottom). The annealing times are varied (see legends).}
    \label{fig:minor_embedded_GSP_QUBOs}
\end{figure}

Figure~\ref{fig:minor_embedded_GSP_QUBOs} shows ground state success probability (GSP) measurements as a function of the chain strength, computed for the $5$ fixed QUBO instances on the four D-Wave annealers of Table~\ref{tab:hardware_summary}. Each subplot additionally showcases the behavior for different annealing times. The figure highlights several observations. First, the \texttt{DW\_2000Q\_6} device seems to achieve a considerably lower GSP than the other devices, followed by \texttt{Advantage\_system4.1} and \texttt{Advantage\_system6.1}, while \texttt{Advantage2\_prototype1.1} achieves highest GSP across the instances. Second, the anneal times do influence the solution quality throughout all instances, with longer annealing times usually resulting in an increased solution quality. Third, although the $5$ QUBO instances were generated with the same parameters, there seems to be a considerable range in difficulty, with the instances in the left columns being harder to solve than the ones in the rightmost columns. 

\section{Discussion}
\label{sec:discussion}
This paper proposes a new method, called posiform planting, to generate QUBOs with a unique planted solution. Apart from guaranteeing the uniqueness, posiform planting can be adapted to any arbitrary connectivity structure, meaning that it allows one to generate tailored QUBO instances whose quadratic couplers fit, for instance, the hardware connectivity of modern quantum annealers. Therefore, posiform planting allows one to efficiently generate QUBO instances with thousands of variables and a unique planted solution. Posiform planting also generates QUBOs that have linear terms, a property that not all of existing planted solution methods have.

Interestingly, our construction shows that, for the generation of a QUBO with a planted solution, the coefficients in a posiform representation do not matter and can be freely chosen (as long as they are positive as required by definition of a posiform). Since the choice of the posiform coefficients does not impact the planted solution or its uniqueness, posiform planted allows for an efficient generation of a set of QUBO instances having the same planted solution. Posiform planting also allows for an arbitrary bitstring to be chosen as the planted solution.

Posiform planting can be used to verify whether good classical heuristic algorithms, such as simulated annealing, are able to find the single optimal solution for extremely large QUBOs. This not only applies to classical algorithms, but also other emerging computing technologies such as spiking neuromorphic computing \cite{7966350, Mniszewski} or the hybrid quantum-classical gate model algorithm QAOA \cite{Hadfield_2019, farhi2014quantum}. We experimentally demonstrated that four D-Wave quantum annealers, with a total of $3$ different classes of hardware graphs, can sample the unique planted solution for hardware native QUBO problems that use the entire hardware chip. Since we scaled our instances to the maximal size that can be embedded on D-Wave, the current hardware limitations (of maximally 5627 qubits on D-Wave Advantage) somewhat limit us from scaling our instances to sizes where D-Wave starts to struggle.

Posiform planting generates QUBOs of a special form in order to guarantee the uniqueness of the planted solution. To be precise, all QUBOs generated by posiform planting have the property that when converted to a posiform representation, they are solvable (meaning they attain a value of zero). However, not all QUBOs have this property. Nevertheless, posiform planting is complete in the sense that it can generate any QUBO whose posiform representation is solvable.

The paper leaves scope for further avenues of research. Most importantly, it remains to investigate if posiform planting allows one to tune the difficulty of the QUBO problems, for instance via the choice of the posiform coefficients (which can be tuned without constraints other than being positive). Another topic for future research is to be able to vary the ground state degeneracy of posiform planted QUBOs, if there are specific use cases where obtaining a QUBO with a specific number of ground states would be advantageous.

Finally, posiform planting can enhance an existing planted solution method, denoted as $M$, to guarantee the uniqueness of the planted solution. For instance, given the desired solution $x^*$ to be planted, we first use $M$ to produce a QUBO $Q_1$ with the planted solution $x^*$, which may be non-unique. Subsequently, leveraging posiform planting, we generate a QUBO $Q_2$ that ensures $x^*$ is a unique optimal solution. By forming the linear combination $Q_{\text{new}} = \alpha_1 Q_1 + \alpha_2 Q_2$ with $\alpha_1, \alpha_2 > 0$, we obtain a problem with a unique solution $x^*$, while potentially preserving any desired properties of $M$.

\section*{Acknowledgments}
\label{sec:acknowledgments}
This work was supported by the U.S.\ Department of Energy through the Los Alamos National Laboratory. Los Alamos National Laboratory is operated by Triad National Security, LLC, for the National Nuclear Security Administration of U.S.\ Department of Energy (with Contract No.~89233218CNA000001). Research presented in this article was supported by the NNSA's Advanced Simulation and Computing Beyond Moore's Law Program at Los Alamos National Laboratory, and was supported by the Laboratory Directed Research and Development program of Los Alamos National Laboratory under project 20210114ER. This research used resources provided by the Los Alamos National Laboratory Institutional Computing Program, which is supported by the U.S. Department of Energy National Nuclear Security Administration under Contract No.~89233218CNA000001. This research used resources provided by the Darwin testbed at Los Alamos National Laboratory (LANL), which is funded by the Computational Systems and Software Environments subprogram of LANL's Advanced Simulation and Computing program (NNSA/DOE). This work has been assigned LANL technical report number LA-UR-23-27170.

The work of Hristo Djidjev was supported by grant number KP-06-DB/1 of the Bulgarian National Science Fund. 

Funding for Georg Hahn was provided through Cure Alzheimer's Fund, the National Institutes of Health [1R01 AI 154470-01; 2U01 HG 008685; R21 HD 095228 008976; U01 HL 089856; U01 HL 089897; P01 HL 120839; P01 HL 132825; 2U01 HG 008685; R21 HD 095228, P01HL132825], the National Science Foundation [NSF PHY 2033046; NSF GRFP 1745302], and a NIH Center grant [P30-ES002109].

\appendix

\section{Hardware native QUBO structures}
\label{sec:appendix_QUBO_hardware}
\begin{figure}[h]
    \centering
    \includegraphics[width=0.49\textwidth]{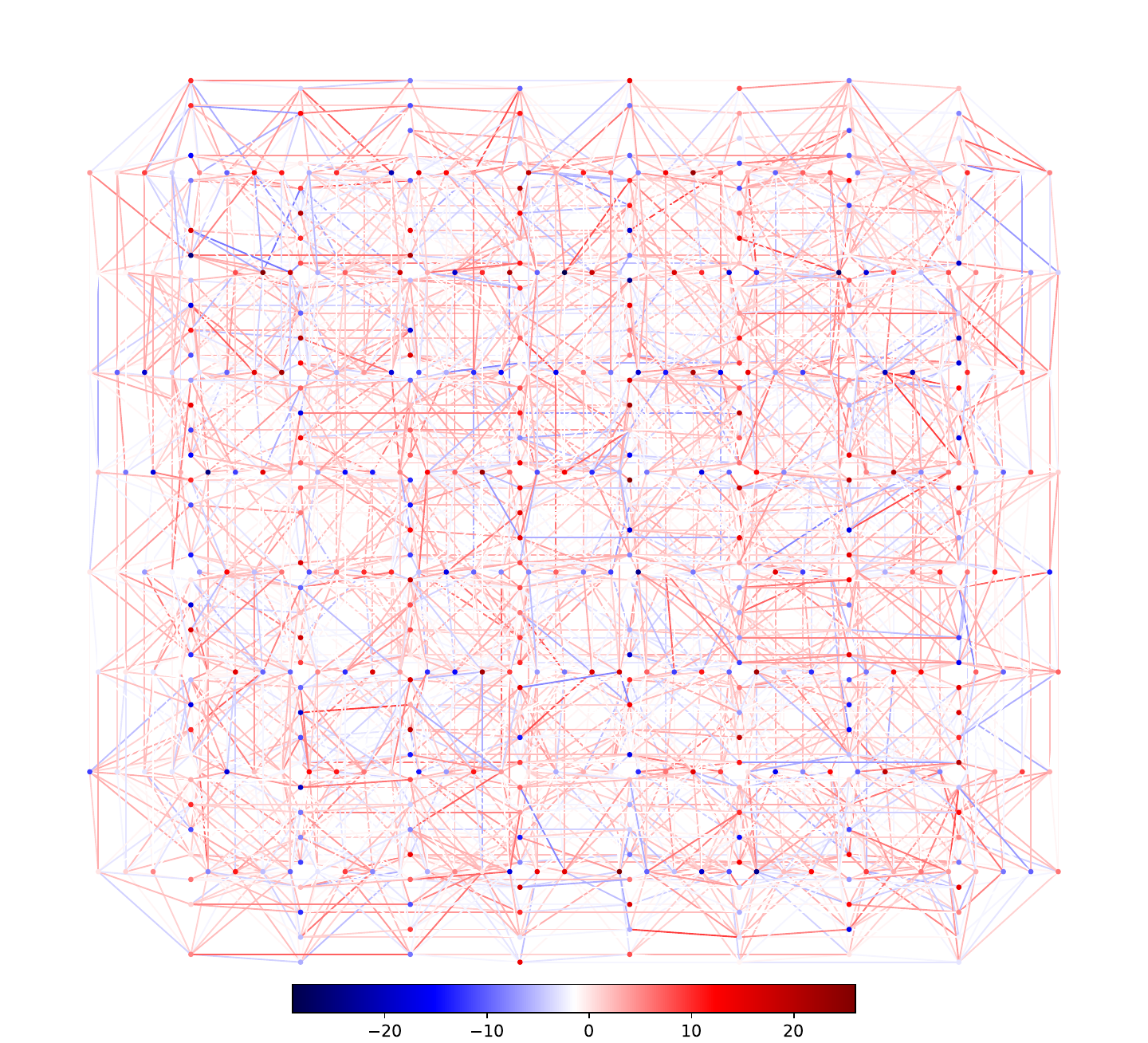}
    \includegraphics[width=0.49\textwidth]{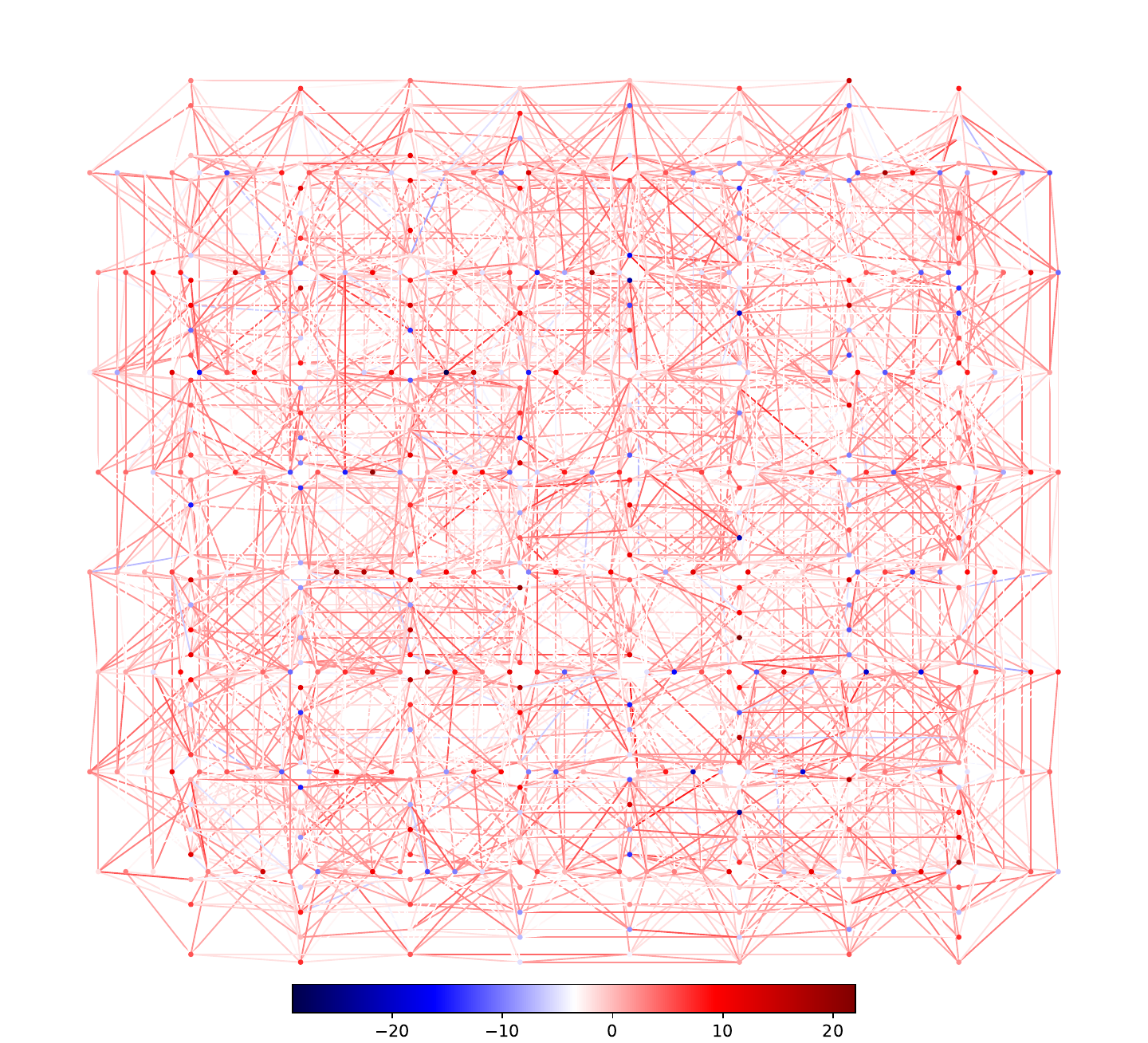}
    \includegraphics[width=0.49\textwidth]{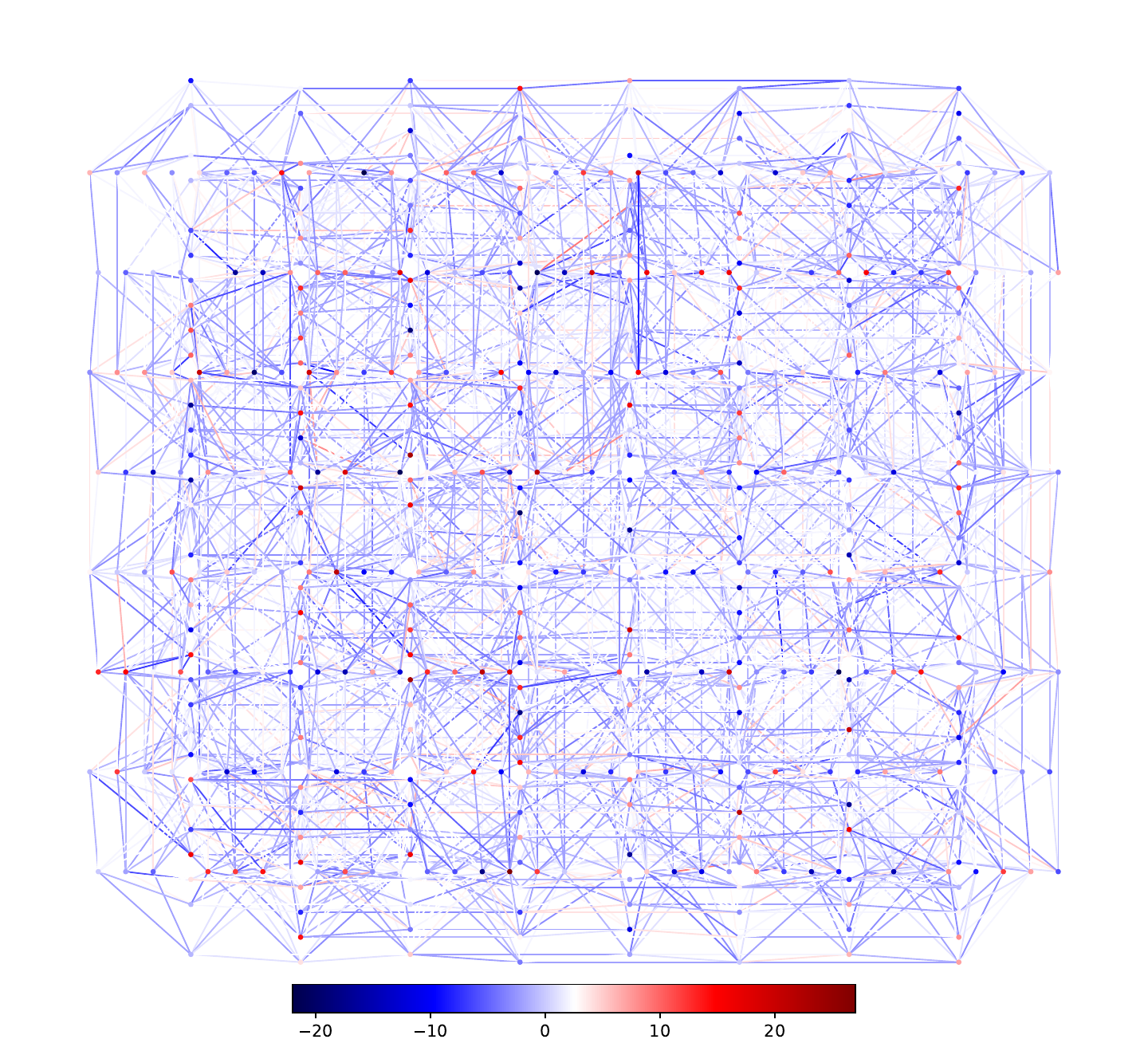}
    \includegraphics[width=0.49\textwidth]{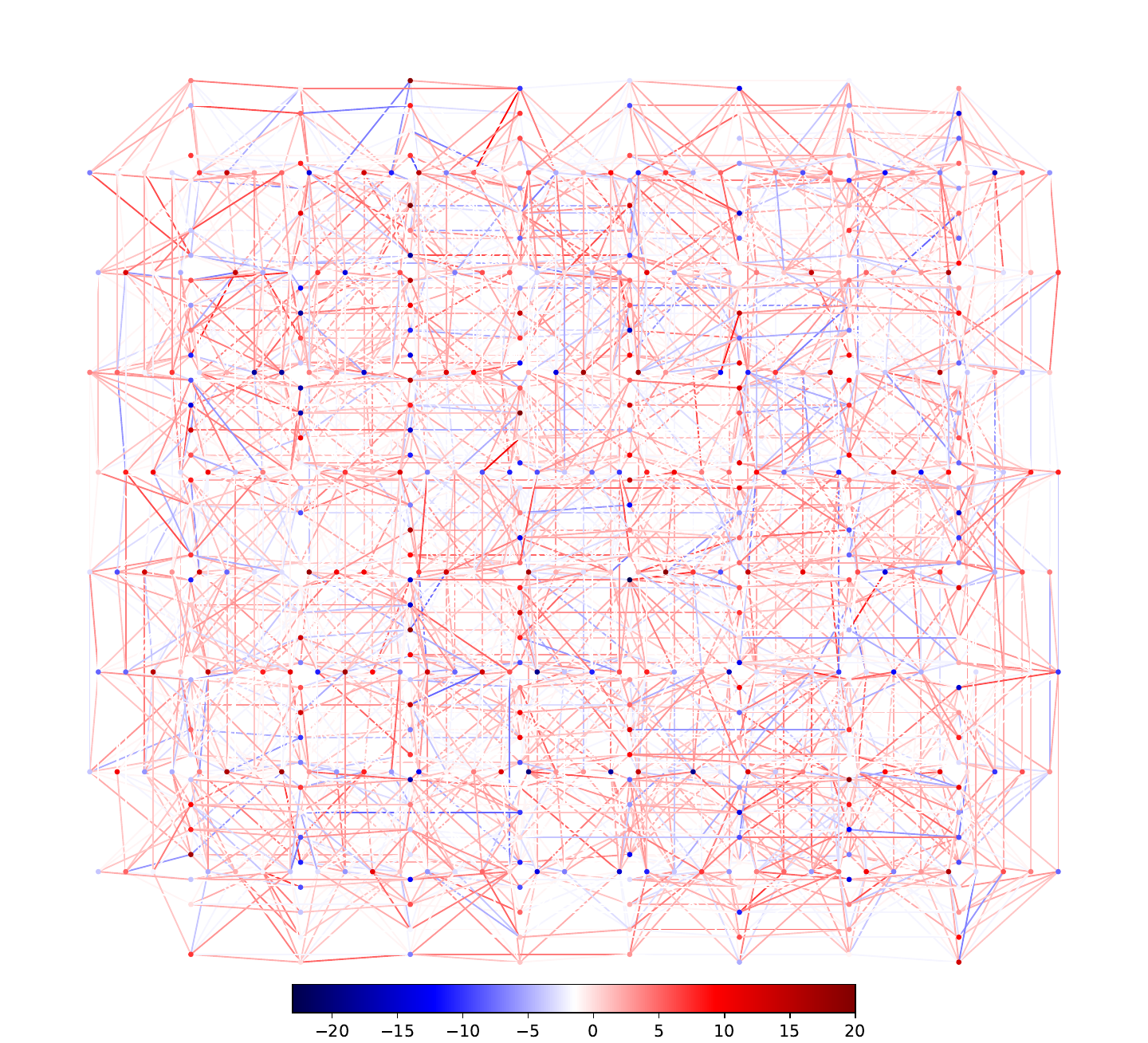}
    \caption{Four example \texttt{Advantage2\_prototype1.1} hardware native posiform planted QUBO coefficient plots. Some hardware couplers are set to $0$ in these QUBOs (these edges are simply drawn as white in the hardware diagrams). Coefficients are encoded in the colormaps, shown below each plot. }
    \label{fig:QUBO_coefficients_native_QUBO_Zephyr_appendix}
\end{figure}

\begin{figure}[h]
    \centering
    \includegraphics[width=0.49\textwidth]{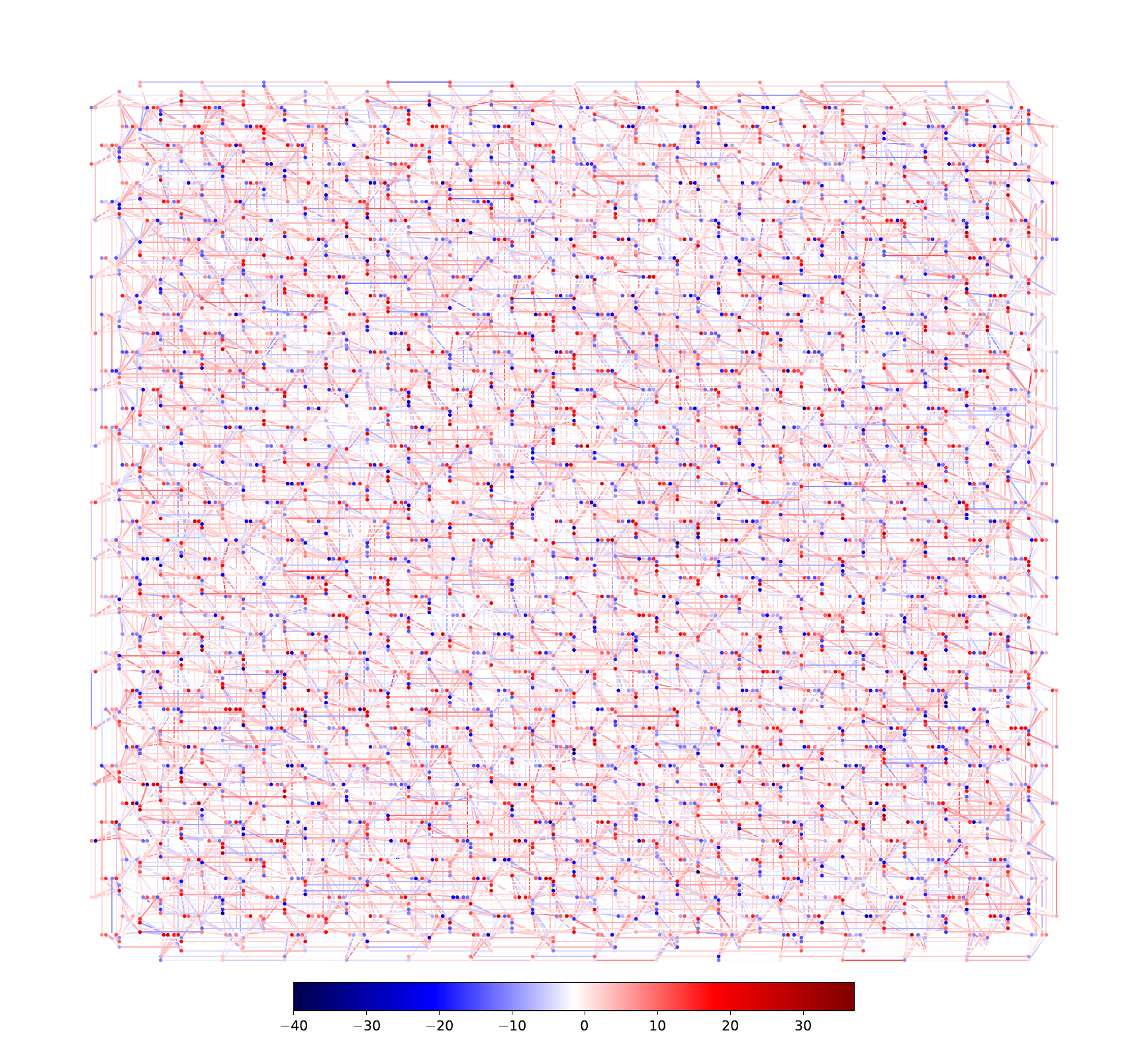}
    \includegraphics[width=0.49\textwidth]{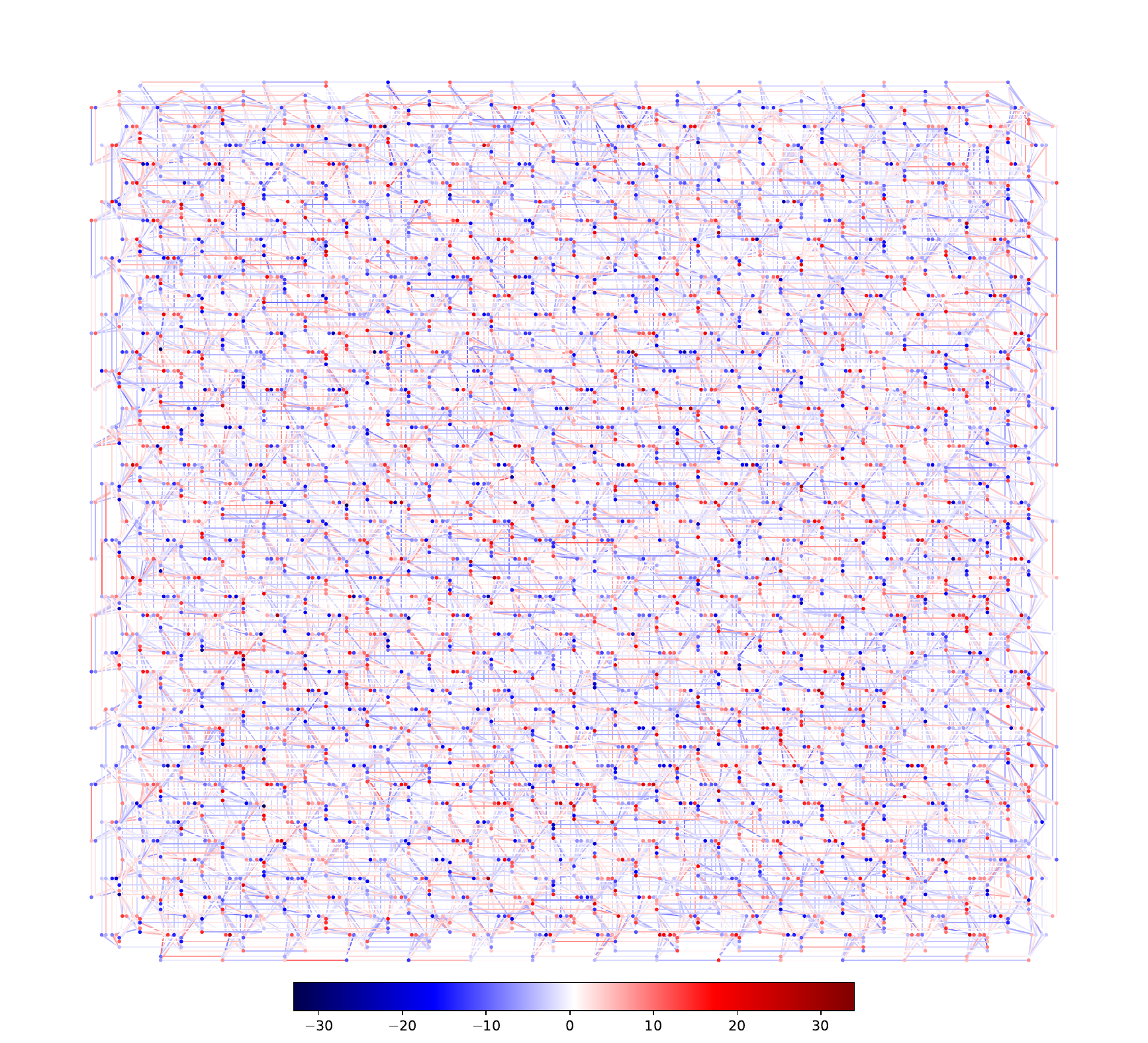}
    \includegraphics[width=0.49\textwidth]{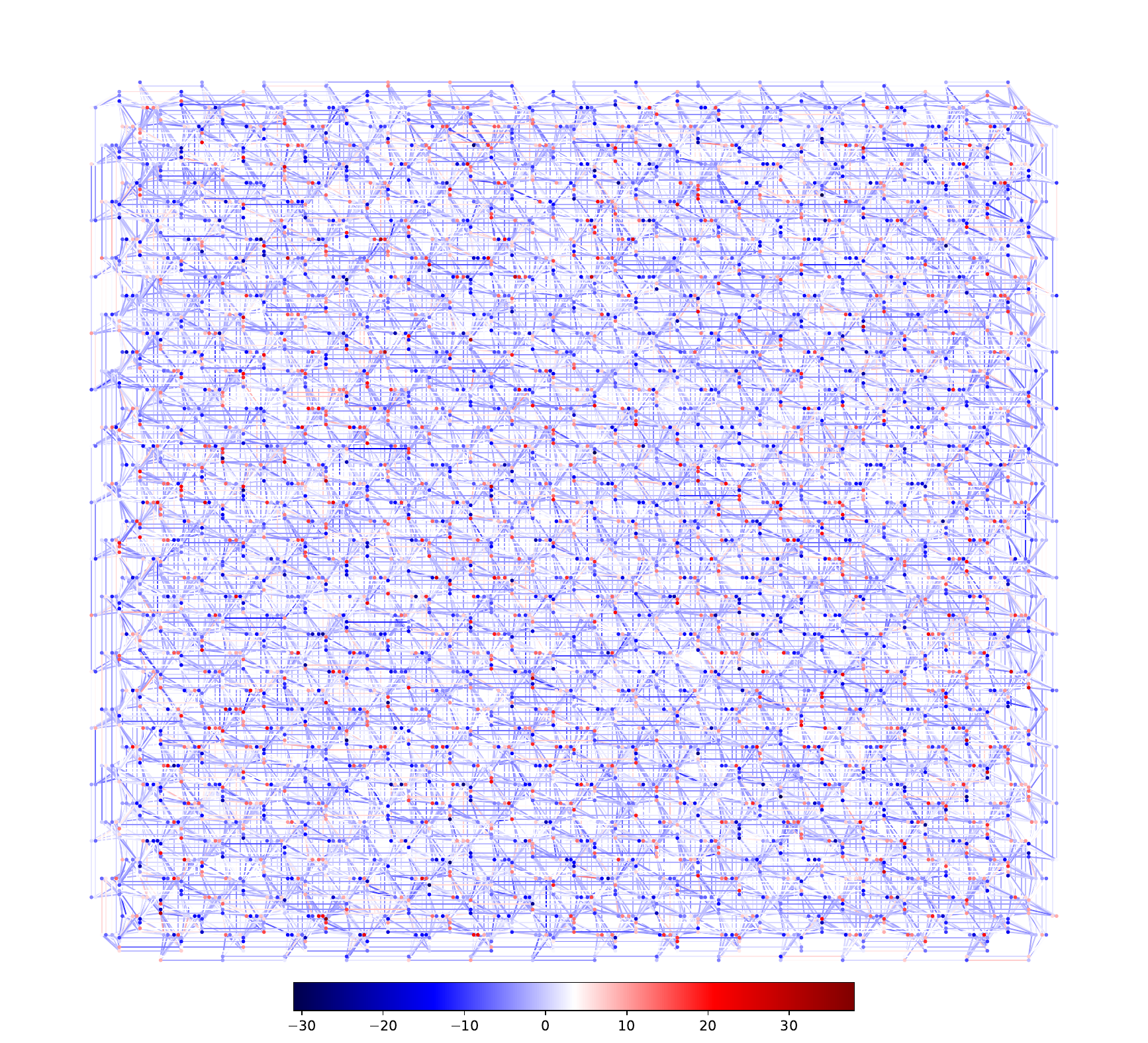}
    \includegraphics[width=0.49\textwidth]{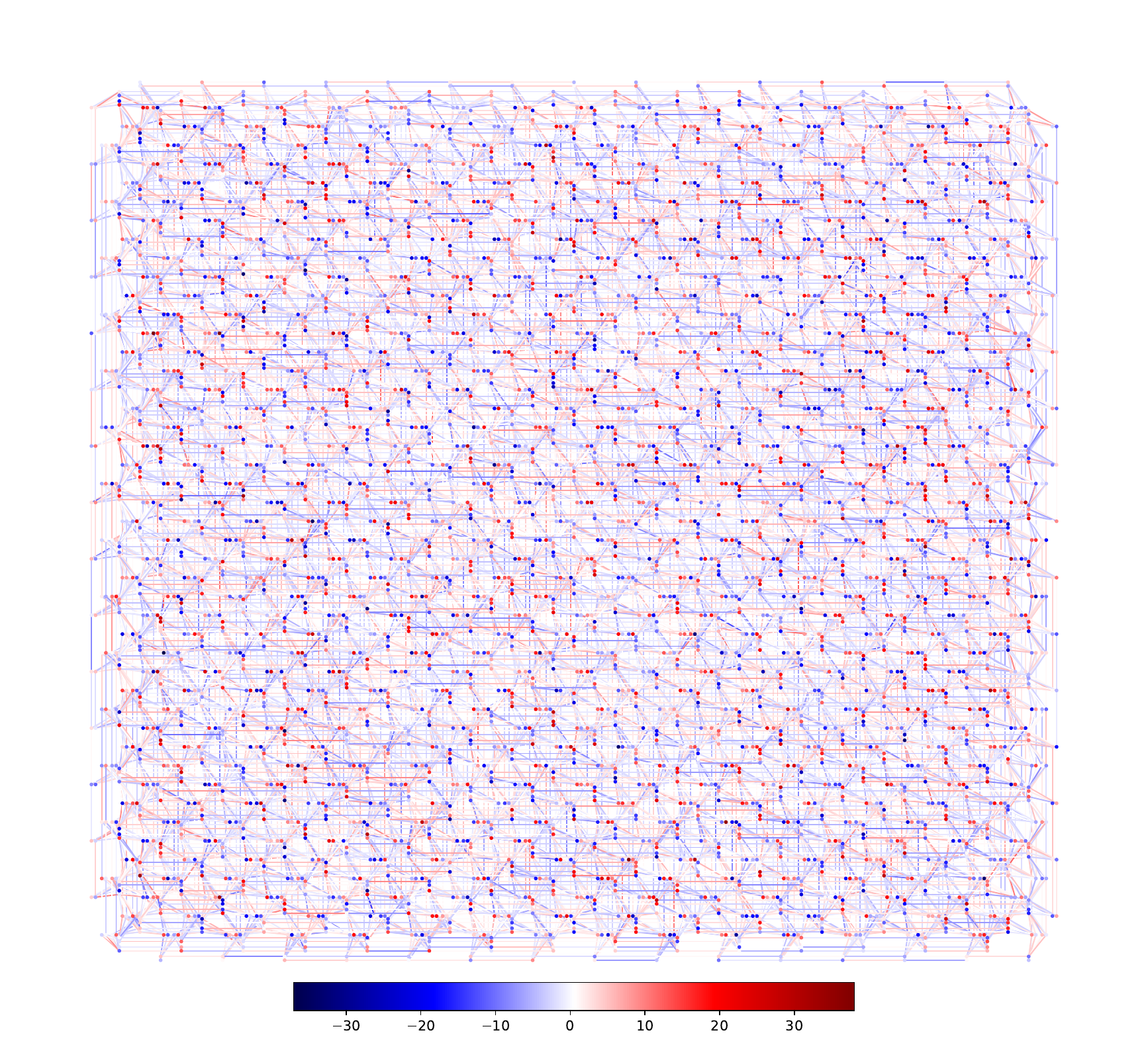}
    \caption{Four example \texttt{Advantage\_system6.1} (top row), \texttt{Advantage\_system4.1} (bottom row) hardware native posiform planted QUBO coefficient plots. Some hardware couplers are set to $0$ in these QUBOs (these edges are simply drawn as white in the hardware diagrams). Coefficients are encoded in the colormaps, shown below each plot. }
    \label{fig:QUBO_coefficients_native_QUBO_Pegasus_appendix}
\end{figure}

\begin{figure}[h]
    \centering
    \includegraphics[width=0.49\textwidth]{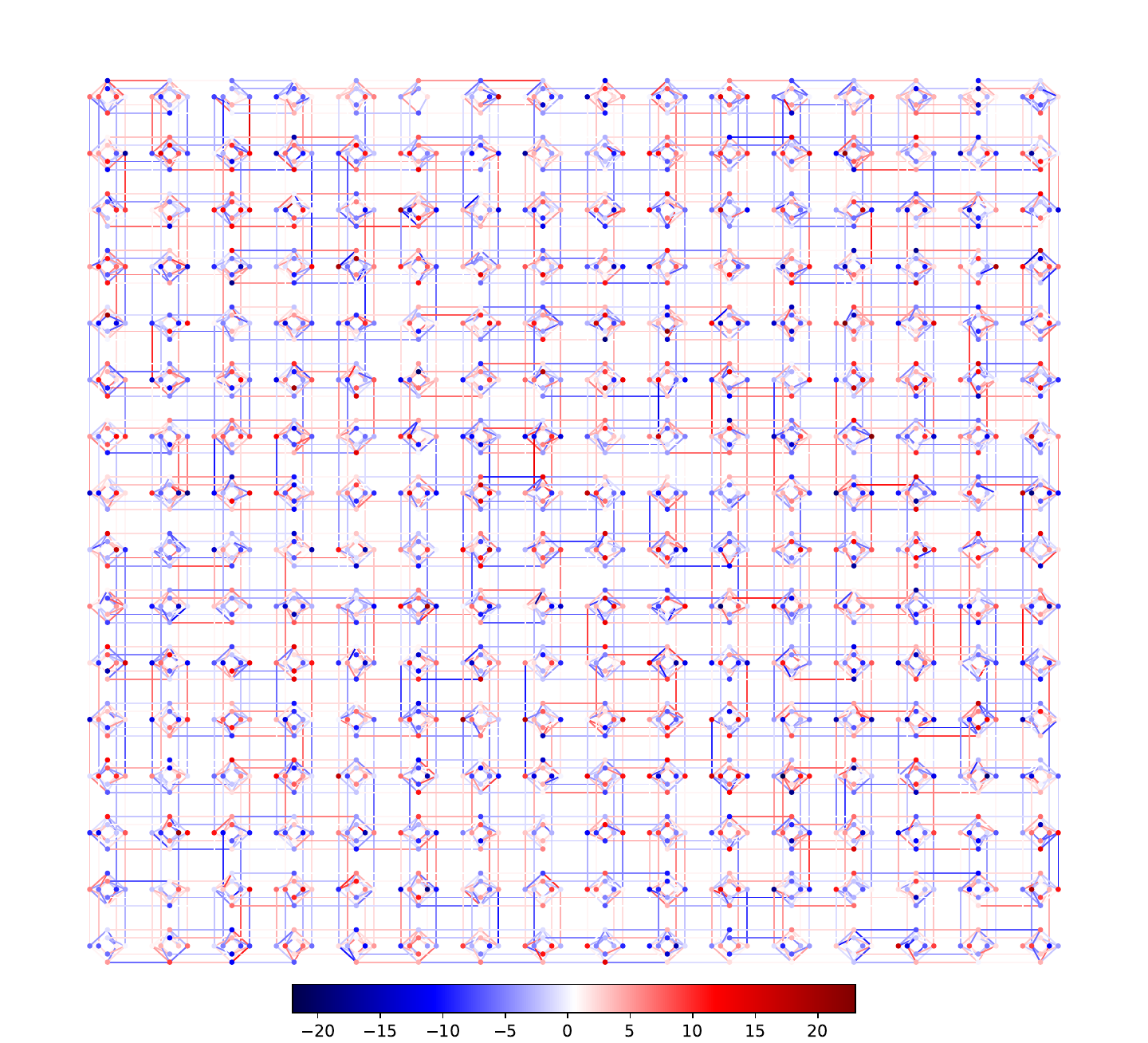}
    \includegraphics[width=0.49\textwidth]{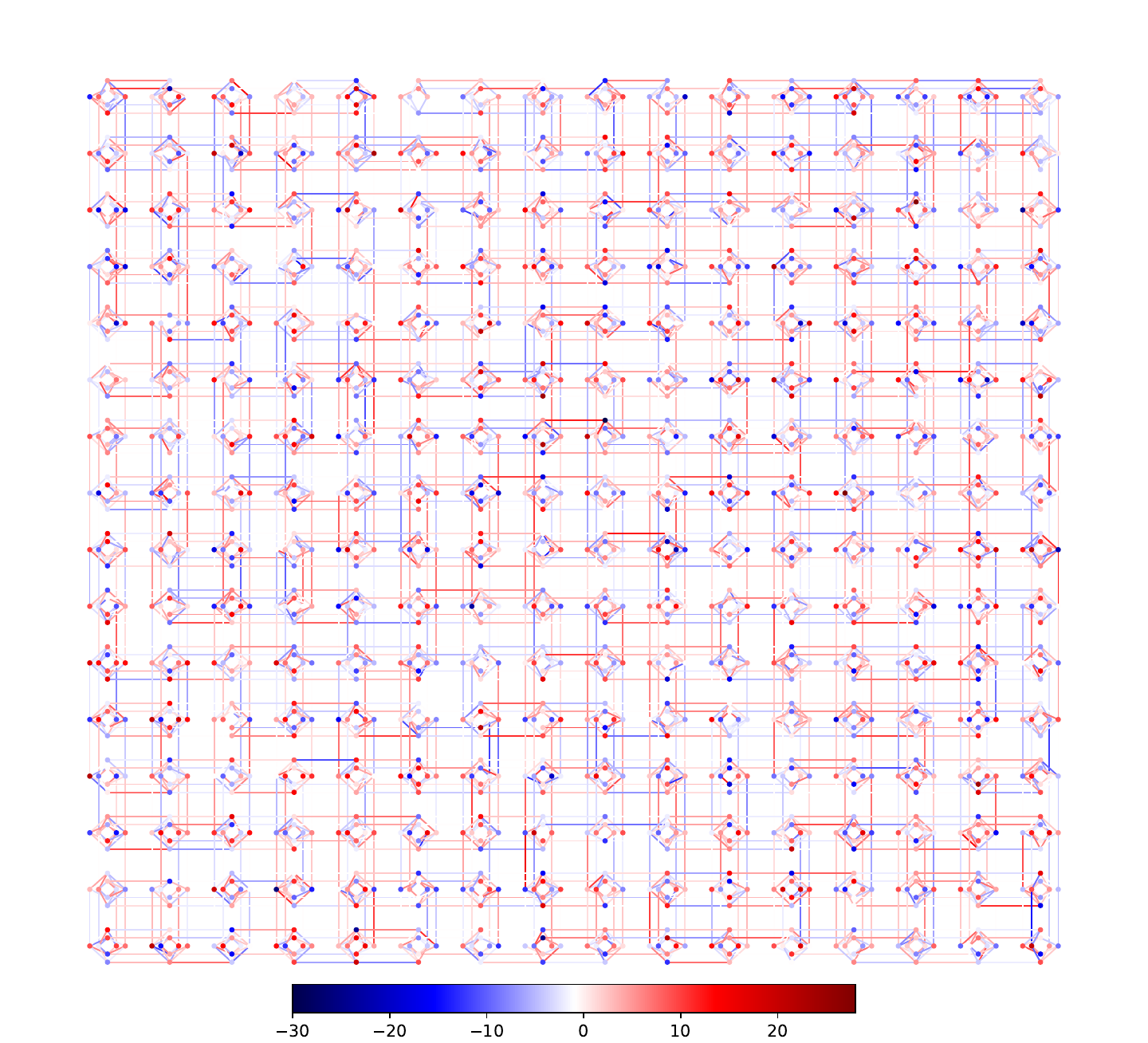}
    \includegraphics[width=0.49\textwidth]{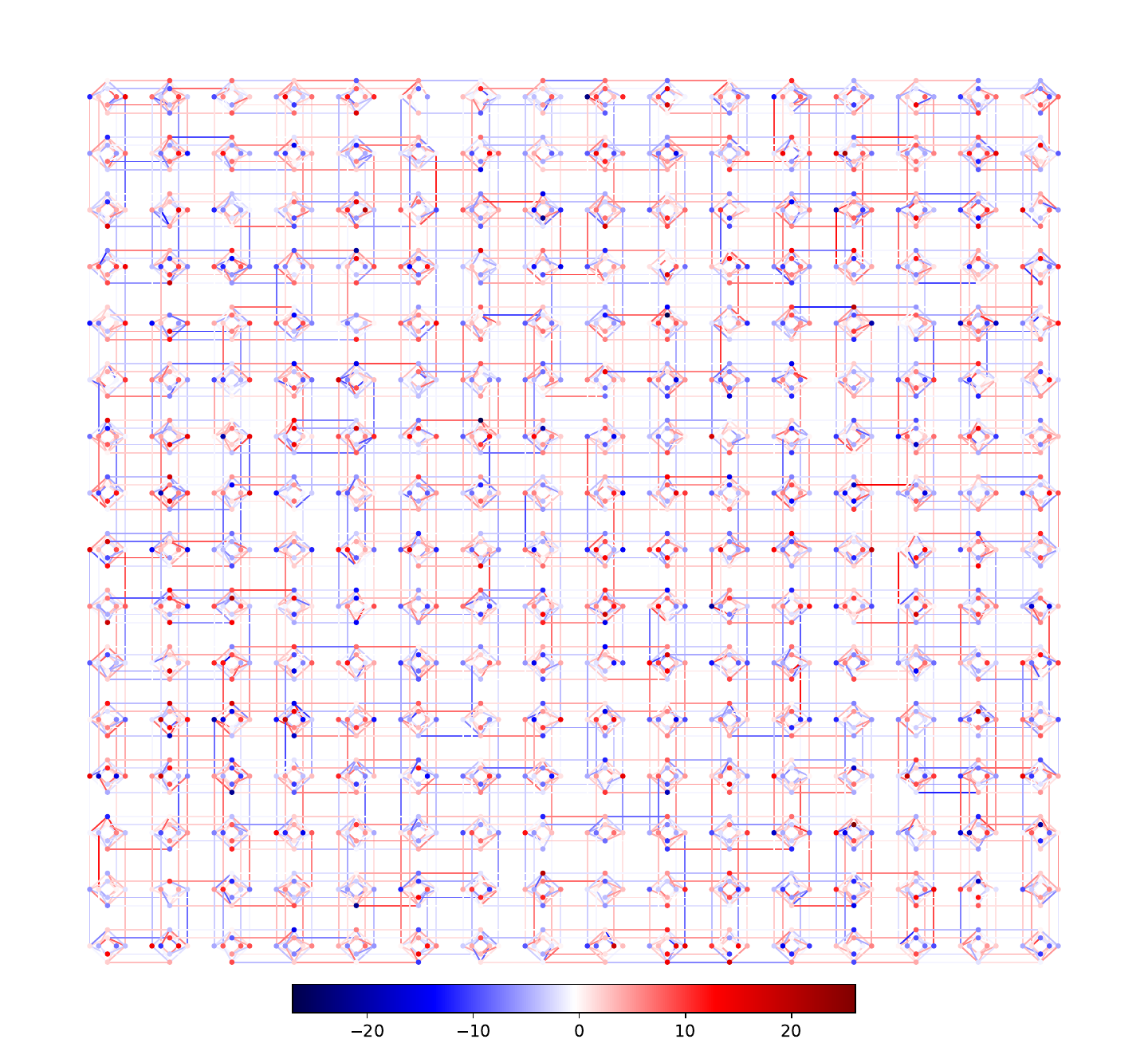}
    \includegraphics[width=0.49\textwidth]{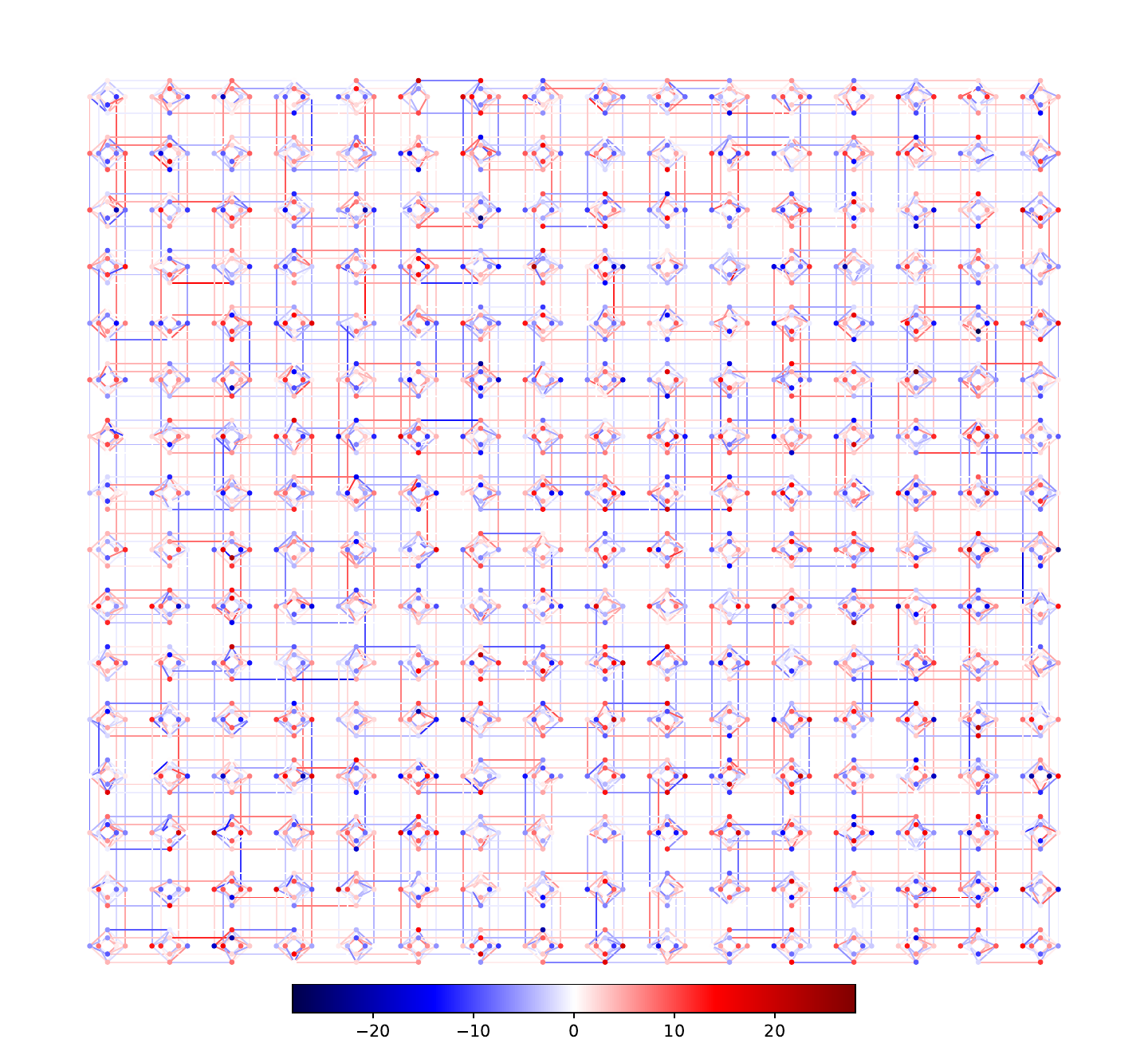}
    \caption{Four example \texttt{DW\_2000Q\_6} hardware native posiform planted QUBO coefficient plots. Some of the hardware couplers are set to $0$ in these QUBOs (these edges are simply drawn as white in the hardware diagrams). Coefficients are encoded in the colormaps, shown below each plot. }
    \label{fig:QUBO_coefficients_native_QUBO_Chimera_appendix}
\end{figure}

An interesting property of the posiform solution planting algorithm is that the resulting QUBOs indeed vary substantially. In particular, for fixed hardware graphs, the QUBO coefficients can vary significantly, and moreover there can be clear coefficient preferences where some posiform planted QUBOs are more positively or negatively weighted. In order to illustrate this visually, Figures~\ref{fig:QUBO_coefficients_native_QUBO_Zephyr_appendix}, \ref{fig:QUBO_coefficients_native_QUBO_Pegasus_appendix}, \ref{fig:QUBO_coefficients_native_QUBO_Chimera_appendix} plot the hardware native QUBOs, including the coefficients encoded using colors. Figures~\ref{fig:QUBO_coefficients_native_QUBO_Zephyr_appendix}, \ref{fig:QUBO_coefficients_native_QUBO_Pegasus_appendix}, \ref{fig:QUBO_coefficients_native_QUBO_Chimera_appendix} show that the posiform planted QUBOs can have coefficients which vary up to between approximately $-40$ and $+40$, depending on the hardware graph.

Because the posiform planting algorithm can terminate without using all of the available underlying connectivity graph, the generated QUBOs do not use all of the available hardware couplers (see Table \ref{tab:hardware_summary}), and this can be seen in Figures~\ref{fig:QUBO_coefficients_native_QUBO_Zephyr_appendix}, \ref{fig:QUBO_coefficients_native_QUBO_Pegasus_appendix}, \ref{fig:QUBO_coefficients_native_QUBO_Chimera_appendix}. On average, across the $100$ randomly generated hardware native QUBOs, $\approx 5330$ couplers were used on the \texttt{DW\_2000Q\_6} instances, $\approx 3503$ couplers were used on the \texttt{Advantage2\_prototype1.1}, $\approx 34028$ couplers were used on the \texttt{Advantage\_system4.1} instances, $\approx 34093$ couplers were used on the \texttt{Advantage\_system6.1} instances.

\section{N=52 Minor Embeddings}
\begin{figure}[h]
    \centering
    \includegraphics[width=0.24\textwidth]{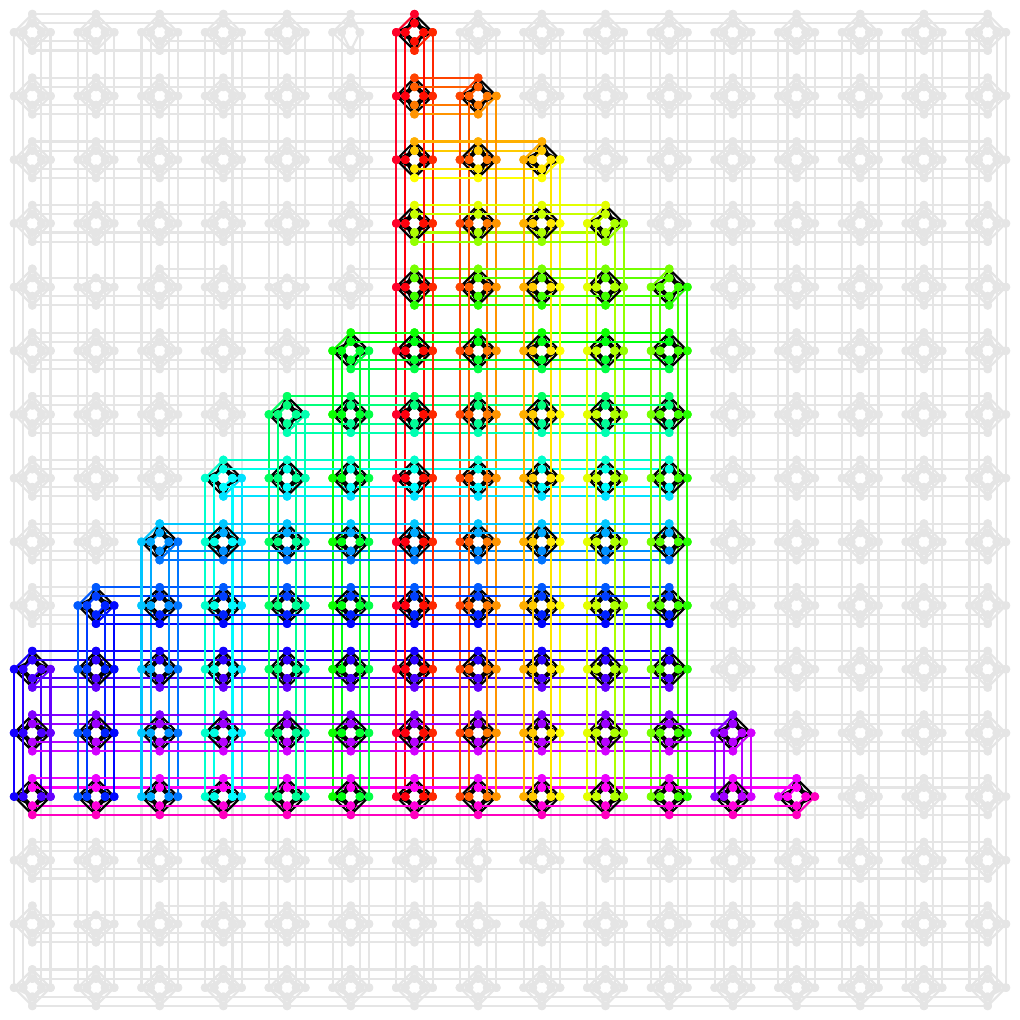}
    \includegraphics[width=0.24\textwidth]{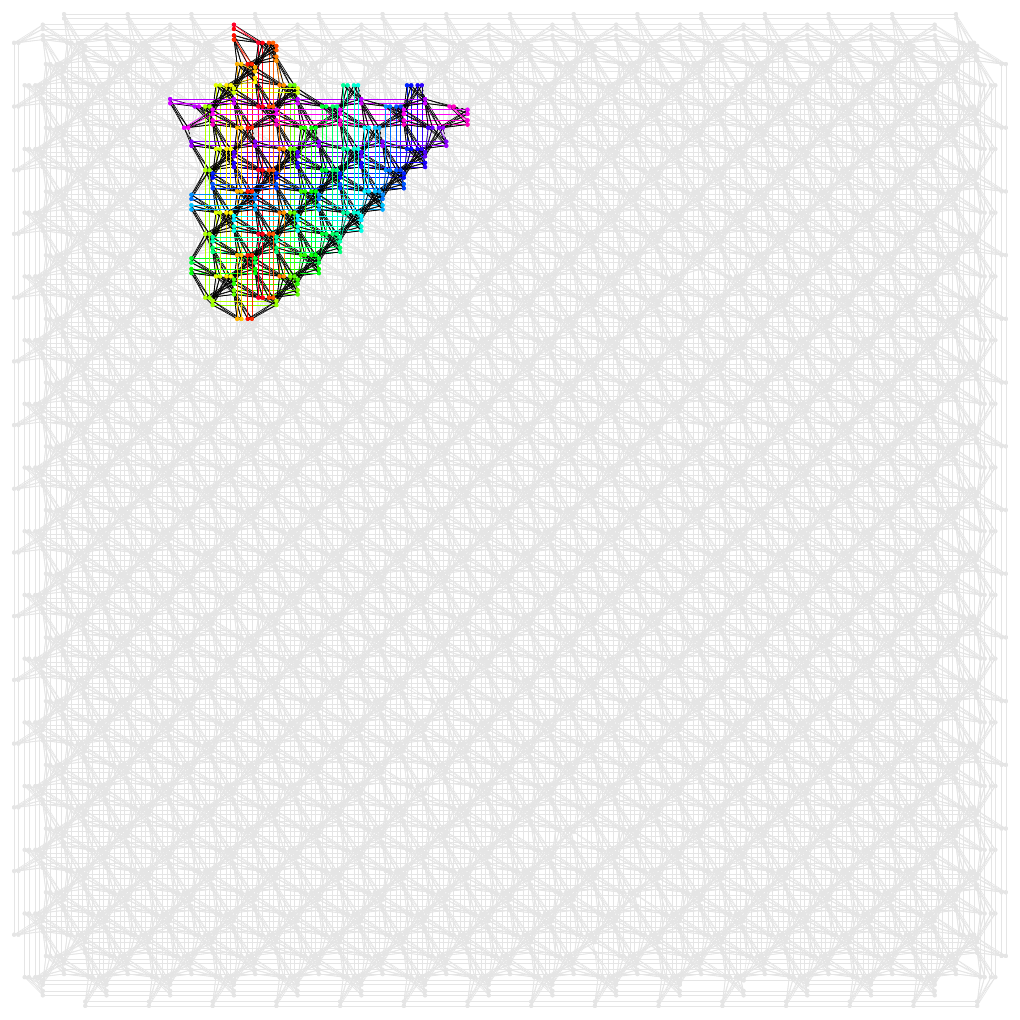}
    \includegraphics[width=0.24\textwidth]{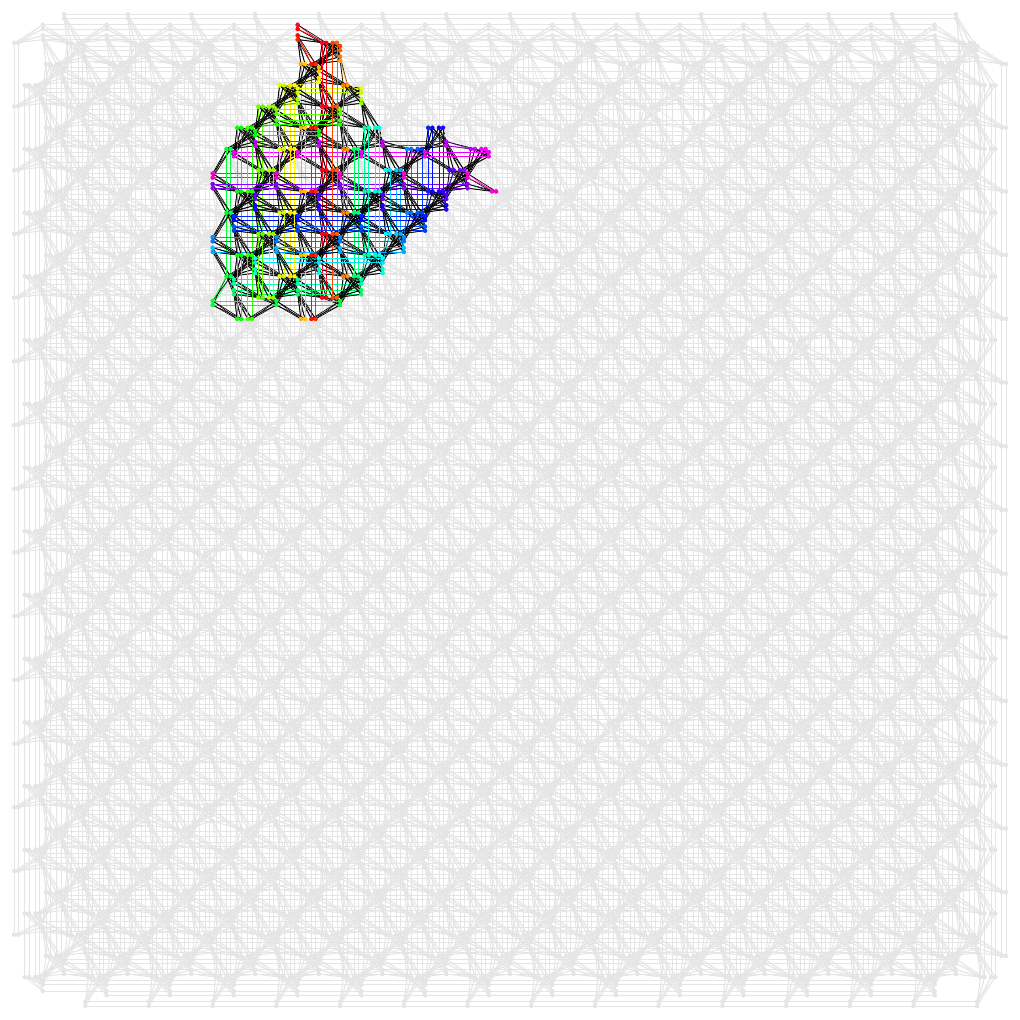}
    \includegraphics[width=0.24\textwidth]{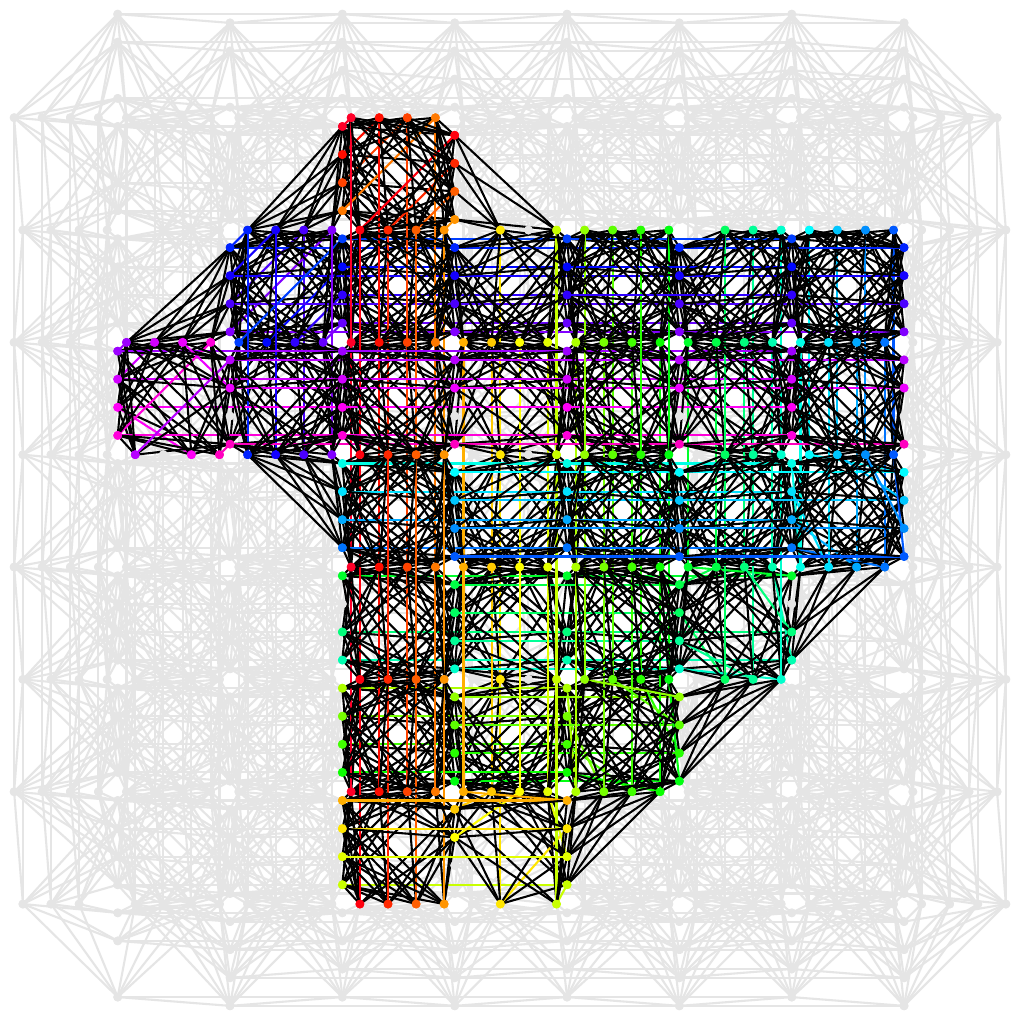}
    \caption{$N=52$ all-to-all minor embeddings across the four QPU hardware graphs. Each chain is representing a logical variable, and each chain in this diagram is colored a consistent color when drawn on the hardware. The chains of qubits have couplers linking them to all other chains, meaning that arbitrarily structured graph problems can be embedded onto these minor embeddings. It is clear that the Zephyr graph minor embedding (right) is much more densely connected, therefore requiring smaller chain lengths, compared to the Chimera graph embedding (left). These structured minor embeddings are specifically generated to have relatively uniform chain lengths, making the quantum annealing solutions of the minor embedded problems better than random minor embeddings with highly variable chain lengths. These diagrams show the minor embeddings onto the hardware graphs of \texttt{DW\_2000Q\_6} (left), \texttt{Advantage\_system4.1} (middle-left), \texttt{Advantage\_system6.1} (middle-right), and \texttt{Advantage2\_prototype1.1} (right). These hardware graphs are logical Chimera $C_{16}$ (left), Pegasus $P_{16}$ (middle-left and middle-right), and Zephyr $Z_{4}$ graphs each with some hardware defects leading to missing qubits and couplers compared to the logical graph structure. }
    \label{fig:52_minor_embedding}
\end{figure}

Figure~\ref{fig:52_minor_embedding} shows the exact $N=52$ all-to-all connectivity minor embeddings used in Section~\ref{sec:results_minor_embedded}. 

\clearpage

\setlength\bibitemsep{0pt}
\printbibliography

\end{document}